\documentclass[final,5p,times,twocolumn]{elsarticle}

\usepackage{amssymb}
\usepackage{amsmath}
\usepackage{amsthm}
\usepackage{amsfonts}
\usepackage{color}
\usepackage{epstopdf}

\usepackage{subcaption}
\usepackage{graphicx}

\newcommand{\tcr}{\textcolor{black}}
\newcommand{\tcb}{\textcolor{black}}

\DeclareMathOperator{\sign}{\operatorname{sign}}

\usepackage[english]{babel}

\journal{Journal of Neuroscience Methods}

\begin{document}

\begin{frontmatter}

\author[ISC]{Thomas Kreuz}
\ead{thomas.kreuz@cnr.it}
\author[UniFi]{Federico Senocrate}
\author[UB]{Gloria Cecchini}
\author[UniFi,LENS]{Curzio Checcucci}
\author[LENS,PISA]{Anna Letizia Allegra Mascaro}
\author[UniFi,LENS,PISA]{Emilia Conti}
\author[UniFi,LENS]{Alessandro Scaglione}
\author[UniFi,LENS,INO]{Francesco Saverio Pavone}


\address[ISC]{Institute for Complex Systems (ISC), National Research Council (CNR), Sesto Fiorentino, Italy}
\address[UniFi]{Department of Physics and Astronomy, University of Florence, Sesto Fiorentino, Italy}
\address[UB]{Department of Mathematics and Computer Science, University of Barcelona, Barcelona, Spain}
\address[LENS]{European Laboratory for Non-linear Spectroscopy (LENS), University of Florence, Sesto Fiorentino, Italy}
\address[PISA]{Neuroscience Institute, National Research Council (CNR), Pisa, Italy}
\address[INO]{National Institute of Optics (INO), National Research Council (CNR), Sesto Fiorentino, Italy}

\title{Latency correction in sparse neuronal spike trains}

\begin{abstract}
\textit{Background:} In neurophysiological data, latency refers to a global shift of spikes from one spike train to the next, either caused by response onset fluctuations or by finite propagation speed. Such systematic shifts in spike timing lead to a spurious decrease in synchrony which needs to be corrected.

\noindent \textit{New Method:} We propose a new algorithm of multivariate latency correction suitable for sparse data for which the relevant information is not primarily in the rate but in the timing of each individual spike. The algorithm is designed to correct systematic delays while maintaining all other kinds of noisy disturbances. It consists of two steps, spike matching and distance minimization between the matched spikes using simulated annealing.

\noindent \textit{Results:} We show its effectiveness on simulated and real data: cortical propagation patterns recorded via calcium imaging from mice before and after stroke. Using simulations of these data we also establish criteria that can be evaluated beforehand in order to anticipate whether our algorithm is likely to yield a considerable improvement for a given dataset.

\noindent \textit{Comparison with Existing Method(s):} Existing methods of latency correction rely on adjusting peaks in rate profiles, an approach that is not feasible for spike trains with low firing in which the timing of individual spikes contains essential information.

\noindent \textit{Conclusions:} For any given dataset the criterion for applicability of the algorithm can be evaluated quickly and in case of a positive outcome the latency correction can be applied easily since the source codes of the algorithm are publicly available.
\end{abstract}


\end{frontmatter}

%
\section{Introduction} \label{s:Intro}
%
%
%
%
%
%
%

Measuring the degree of synchrony within a set of spike trains is a common task in two major scenarios. 

In the first scenario spike trains are recorded in \textit{successive} time windows from only \textit{one neuron}. In order to allow for a meaningful alignment of these time windows there has to be a common temporal reference point which is typically some kind of external trigger event such as the onset of a stimulus. If the stimulus is always the same, the issue under consideration is the reliability of individual neurons \cite{Mainen95, Tiesinga08}, while different stimuli are used to find the features of the response that provide the optimal discrimination within the context of neuronal coding \cite{Victor05, QuianQuiroga13}.

In the second scenario spike trains are recorded \textit{simultaneously} from a \textit{population of neurons} \cite{Gerstein01, Brown04}. In this scenario spikes emitted at the same time are truly `synchronous' (Greek: `occurring at the same time'). Typical applications for such data are the multi-channel recordings of various neuronal circuits of the brain \cite{Tiesinga08, Shlens08} and the analysis of spiking activity propagation in neuronal networks \cite{Buzsaki04, Kumar10}. The real data example used in this study also belongs to this scenario, global activation patterns recorded via wide-field calcium imaging in the cortex of mice before and after stroke \cite{mascaro2019combined, Cecchini21}.

Generally speaking, \textit{latency} is the temporal delay of some physical change in the system being observed. In neuroscience latencies have biological reasons and carry a lot of valuable information in themselves that are typically analyzed in a first step using measures of directionality \cite{Kreuz17, Cecchini21}. Once all the information contained in the latencies has been extracted, it is worth investigating also the synchrony of the underlying dynamics. In this context of synchrony estimation latency in the data is not a primary source of information but rather a hindrance that first has to be removed in order to find the true value of synchrony.

Latency and its correction are relevant in both of the scenarios introduced above. In the first `successive single neuron recordings' scenario latency translates into the time lag between the stimulus and the response. Due to various sources of noise \cite{lee2020neural, uzuntarla2012controlling} this lag may vary from trial to trial \cite{lee2016signal}. These variations in onset latency can then lead to a ``spurious" underestimation of synchrony \cite{ermentrout2008reliability, zirkle2021noise}. In order to account for this, a multivariate latency correction has to be performed in which the various trials are realigned before the ``true" synchrony is calculated.
 
In the second `simultaneous population recordings' scenario dealt with here, latency becomes important whenever there is a spatial propagation of activity from one location to another \cite{Kreuz17, Cecchini21}. The question of interest when analyzing synchrony becomes whether and how much the activity changes during the course of the propagation. Again, in order to answer this, the recordings from different locations have to be compared only after the latency caused by the finite propagation speed has been accounted for.

Methods of latency correction proposed in the literature mostly deal with the first scenario and use rate-based estimates of latency. Typically they rely on dynamic rate profiles for each trial that are obtained by convolving the individual spike trains with either a static or a dynamic kernel \cite{Nawrot03, Schneider06}. The resulting peaks are then realigned, e.g. by maximizing the total pairwise correlation \cite{Nawrot03}. This is done under the underlying assumptions that the rate is the most important feature of the response, that the number and density of spikes is high enough to estimate it reliably and, crucially, that the timing of the individual spikes can be neglected. These assumptions hold for a wide variety of real data \cite{walter2009advantages, enoka2017rate} but there are also many datasets in which this is not the case \cite{VanRullen01, Harvey13, Fukushima15} and for which so far no reliable and feasible method of latency correction has been proposed.

Therefore, here we would like to address the complementary problem of latency correction in data in which there are not that many spikes and where the relevant information is not primarily in the rate but in the timing of each individual spike. In doing so, we follow two specific objectives: First we would like to propose a latency correction algorithm that works not only with rather clean simulated data but functions also with experimental data that typically contain disturbances such as unreliability (missing spikes), jitter (noisy spike shifts), and background noise (extra spikes). The algorithm consists of two parts, matching pairs of spikes over all the spike trains followed by an optimization procedure (simulated annealing) that minimizes the distances between the matched spikes. The second objective is to define the limits for which the algorithm works well, e.g. to specify whether there are any conditions the dataset needs to fulfill in order to be a good candidate for the application of this method.

The remainder of the article is organized as follows: 

In Section \ref{s:Data} we describe the wide-field calcium imaging datasets that we use to illustrate the effectiveness of the algorithm at work. In the Methods (Section \ref{s:Methods}) we first describe the two basic steps of our latency correction algorithm, matching pairs of spikes and minimizing their average distance via simulated annealing (Section \ref{ss:Methods-Spike-Matching-SimAnn}). Then we define the two quantities SPIKE-Synchronization and Synfire Indicator (based on the measure SPIKE-order) from which we will later derive a well-defined criterion for the improvability of a given dataset (Section \ref{ss:Methods-SPIKE-order-Synfire-Indicator}). The Results (Section \ref{s:Results}) consist of three subsections detailing applications of the new approach to artificially generated datasets (Section \ref{ss:Results-Sim}), to neurophysiological datasets (Section \ref{ss:Results-Exp}) and to simulations of these experimental data (Section \ref{ss:Results-SimExp}). Conclusions are drawn in Section \ref{s:Conclusions}.

%
\section{Data} \label{s:Data}

Here we provide a short overview of the experimental paradigm, the basic recording setup and the data processing that was performed in order to arrive at the rasterplots that were then analyzed in Section \ref{ss:Results-SimExp}. More details can be found in \cite{mascaro2019combined, Cecchini21}. 

The datasets consist of cortical activity obtained by $12 \times 21$ pixel wide-field calcium imaging in mice before and after the induction of a focal stroke via a photothrombotic lesion in the primary motor cortex. The purpose of the recordings was to investigate changing propagation patterns during motor recovery from the functional deficits caused by the stroke. This motor recovery was aided by the M-platform \cite{Spalletti14, Spalletti17}, a robotic system that performs passively actuated contralesional forelimb extension on a slide to trigger active retraction movements that were subsequently rewarded (up to 15 cycles per recording session).

In this article we analyze a total of $260$ recordings (mean duration $217$s, range $68$ to $400$s) from $14$ mice which were divided into three groups according to their rehabilitation paradigm: Control, Robot  and Combined. The healthy \textit{Controls} (3 mice) had no stroke induced but underwent four weeks of motor training. The \textit{Robot group} (5 mice) performed the same physical rehabilitation for four weeks starting five days after the stroke induction. The \textit{Combined group} (6 mice) underwent motor training together with a transient pharmacological inactivation of the controlesional hemisphere.

Each recording session resulted in continuous calcium traces from between $173$ and $252$ pixels (mean number $238$) that were then transformed via a straightforward detection of upwards threshold crossings into the spike trains that together form the rasterplots that are analyzed here. These rasterplots display the global activity propagation events in the cortex that typically correspond to attempted or completed forelimb pull events.

The data that we obtained had already undergone some filtering of background noise. Before applying our multivariate latency correction we sort spike trains from leader to follower by means of the SPIKE-order approach \cite{Kreuz17, Cecchini21}.

All experimental procedures were performed in accordance with directive 2010/63/EU on the protection of animals used for scientific purposes and approved by the Italian Minister of Health, authorization n.183/2016-PR.

%
\section{Methods} \label{s:Methods}

In this Section we describe our approach to multivariate latency correction. The starting point is a set of $N$ spike trains each composed of sequences of spike times ${t_i^{(n)}}, i = 1,...,M_n, n = 1,...,N$ recorded over a certain period of time $[0, T]$ (with $M_n$ denoting the numbers of spikes in spike train $n$).

\begin{figure*}[!ht]
\includegraphics[width=\linewidth]{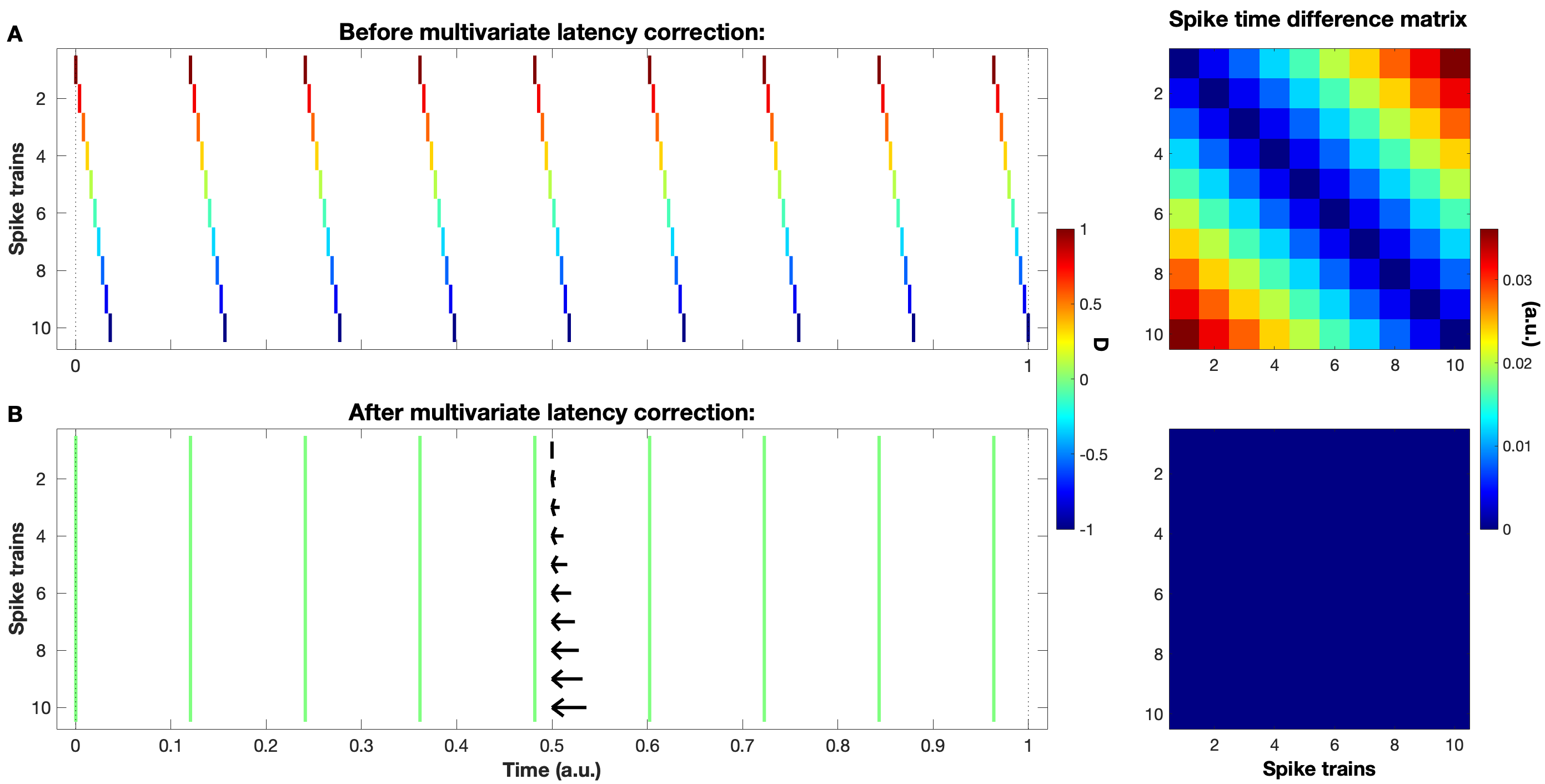}
	\caption{Illustration of a set of artificially created spike trains exhibiting a consistent propagation activity, in this case a perfect regularly spaced synfire chain. Subplot A is before and subplot B is after the multivariate latency correction. The arrows in B indicate for each spike train the shift performed during the latency correction. Spike colors code the asymmetric SPIKE-order $D$ (see Section \ref{ss:Methods-SPIKE-order-Synfire-Indicator}) on a scale from $1$ (red, first leader) to $-1$ (blue, last follower). On the right we show the spike time difference matrices. For this simple example the corrected spike trains are identical and accordingly the spike time difference matrix turns from its perfectly ordered ascension away from the diagonal to zero everywhere.}
\label{F1_Synfire_chain}
\end{figure*}

Since a delay correction for a set of spike trains without any delays is obviously not very reasonable we assume that the set contains spike trains exhibiting a systematic delay between them (later we will define a criterion that tells us if and to what extent this is actually the case). In our preferred `simultaneous population recordings' scenario this corresponds to a consistent propagation of activity from leading to following spike trains. The simplest example, a perfect synfire chain, is shown in the rasterplot of Fig. \ref{F1_Synfire_chain}A. Choosing an arbitrary spike train as reference (here we will always use the first), the task is to find for each of the other $N-1$ spike trains the delay correction that maximizes the similarity (or minimizes the dissimilarity) between all the spike trains. An example of such a solution, in this simple case perfect identity, is shown in Fig. \ref{F1_Synfire_chain}B. 
%
%

Of course in real data any propagation pattern will typically be contaminated by unreliability, jitter, and background noise. And our task is complicated by two more problems:
First, since time is continuous the number of possible solutions to this optimization problem is infinite. Second, in the approaches based on the rate coding assumption \cite{Nawrot03, Schneider06} the individual spike trains are transformed into rate functions with one well-defined maximum each and these maxima can then easily be aligned, but for the kind of sparse data under consideration here calculating rate functions and identifying maxima is so difficult that such a reductive approach is not feasible. 

Instead, our algorithm looks at the temporal relationship between individual spikes and uses an iterative heuristic approach (simulated annealing) to search within the vast space of possible solutions for the optimal one. Since for reasons of computational cost and feasibility we can not calculate in each iteration the overall synchrony among all the spike trains, we use a definition of synchrony based on the measure SPIKE-synchronization that is straightforward to calculate and very easy and efficient to update.

%
\subsection{Spike matching and simulated annealing} \label{ss:Methods-Spike-Matching-SimAnn}

Searching for systematic delays in a spike train set requires a way to determine which spikes should be compared against each other. Under the assumption of sparse data with rather well-defined global events we employ the adaptive coincidence criterion originally introduced for the bivariate measure \textit{event synchronization} \cite{QuianQuiroga02b} and then also used for both the symmetric SPIKE-Synchronization \cite{Kreuz15} and the asymmetric SPIKE-order \cite{Kreuz17}.

This coincidence detection is scale- and parameter-free since the maximum time lag $\tau^{(n,m)}_{ij}$ up to which two spikes $t_i^{(n)}$ and $t_j^{(m)}$ of spike trains $n,m=1,...,N$ are considered to be synchronous is adapted to the local firing rates according to 
\begin{equation} \label{eq:Coincidence-MaxDist}
 \begin{aligned}
    \tau^{(n,m)}_{ij} = \min \{&t_{i+1}^{(n)} - t_i^{(n)}, t_i^{(n)} - t_{i-1}^{(n)},\\
    &t_{j+1}^{(m)} - t_j^{(m)}, t_j^{(m)} - t_{j-1}^{(m)}\}/2.
 \end{aligned}
\end{equation}

For some applications it might be appropriate here to also introduce a maximum coincidence window $\tau_{max}$ \cite{Kreuz17} as a parameter thereby combining the time-scale independent coincident detection with a time-scale dependent upper limit. This way additional knowledge about the data (such as typical signal propagation speed) can be taken into account in order to guarantee that two coincident spikes are really part of the same meaningful event.

Following the derivation of SPIKE-synchronization \cite{Kreuz15}, we then apply the coincidence criterion by defining for each spike $i$ of any spike train $n$ and for each other spike train $m$ a coincidence indicator
\begin{equation} \label{eq:Coincidence-Indicator}
	C_i^{(n,m)}=\begin{cases}
		1 & {\rm if}  \min_j (|t_i^{(n)} - t_j^{(m)}|) < 
\tau_{ij}^{(n,m)} \cr
		0 & {\rm otherwise,}
	\end{cases}
\end{equation}
\noindent which is either one or zero depending on whether this spike is part of a coincidence with a spike of spike train $m$ or not.  This results in an unambiguous spike matching since any spike can at most be coincident with one spike (the nearest one) in the other spike train.

Subsequently, for each spike of every spike train a normalized coincidence counter
\begin{equation} \label{eq:Multi-Counter}
	C_i^{(n)}=\frac{1}{N-1} \sum_{m \neq n} C_i^{(n,m)}
\end{equation}
is obtained by averaging over all $N-1$ bivariate coincidence indicators involving the spike train $n$.

In order to obtain a single multivariate SPIKE-Synchronization profile we pool the coincidence counters of all the spikes of every spike train:
\begin{equation} \label{eq:Multi-Profile}
    	\{C(t_k)\} = \bigcup_n \{C_{i(k)}^{(n(k))} \},
\end{equation}
where we map the spike train indices $n$ and the spike indices $i$ into a global spike index $k$ denoted by the mapping $i(k)$ and $n(k)$. 

With $M = \sum_n M_n$ denoting the total number of spikes in the pooled spike train, the average of this profile
\begin{equation} \label{eq:SPIKE-synchronization}
	C = \begin{cases}
		\frac{1}{M} \sum_{k=1}^M C(t_k) & {\rm if} \  M > 0 \cr
		\ \ \ \ \ \ \ \ \ 1 & {\rm otherwise}
	\end{cases}
\end{equation}
\noindent yields SPIKE-Synchronization, the overall fraction of coincidences. 
It reaches one if and only if each spike in every spike train has one matching spike in all the other spike trains (or if there are no spikes at all), and it attains the value zero if and only if there are no coincidences in any of the spike trains.

For SPIKE-synchronization the only information used is binary: match or no match. Here, for the purpose of latency correction, we go one crucial step further and calculate for each pair of matched spikes the difference between the respective spike times. To do so, for each matched spike in spike train $n$ (all spikes for which $C_i^{(n,m)} = 1$, cf. Eq. \ref{eq:Coincidence-Indicator}) we first identify the matching spike in spike train $m$ as
\begin{equation} \label{eq:Matching-Spike}
	j' = \arg \min_j ( |t_i^{(n)} - t_j^{(m)}| )
\end{equation}
and then calculate their distance as
\begin{equation} \label{eq:Matching-Spike-Time-Difference}
	\delta_i^{(n,m)} = |t_i^{(n)} - t_{j'}^{(m)}|.
\end{equation}
Finally, we obtain the average spike time differences for this spike train pair
\begin{equation} \label{eq:Average-Spike-Time-Difference}
	\delta^{(n,m)} =  \frac{1}{\sum_i C_i^{(n,m)}} \sum_i C_i^{(n,m)} \delta_i^{(n,m)}
\end{equation}
which serves as our best estimate of the latency between these two spike trains. Repeating this procedure for all pairs of spike trains we obtain the symmetric spike time difference matrix $\Delta$. The aim of our multivariate latency correction algorithm is to minimize the mean value of this matrix (or since the matrix is symmetric, the mean value of the upper right tridiagonal part of the matrix $\Delta_<$, i.e., all values for which $n < m$).

In Fig. \ref{F1_Synfire_chain} on the right we show the spike time difference matrix for a perfect synfire chain, before and after the latency correction. In the initial synfire chain, the further apart two spike trains the larger their spike time differences. Accordingly, the values in the spike time difference matrix increase with the distance from the diagonal (which by definition is always zero). In this case the latency correction is very straightforward: a simple shift correction using either the values from the first row (with the first spike train as reference) or the values from the first upper diagonal (the difference between neighboring spike trains) does the trick. In this particular example the shift not only sets the matrix elements that were used in the calculation to zero but also all the other elements of the matrix, as can be seen on the lower right of Fig. \ref{F1_Synfire_chain}. Whenever this is the case, the problem is solved and we stop immediately.

\begin{figure*}[!ht]
\includegraphics[width=\linewidth]{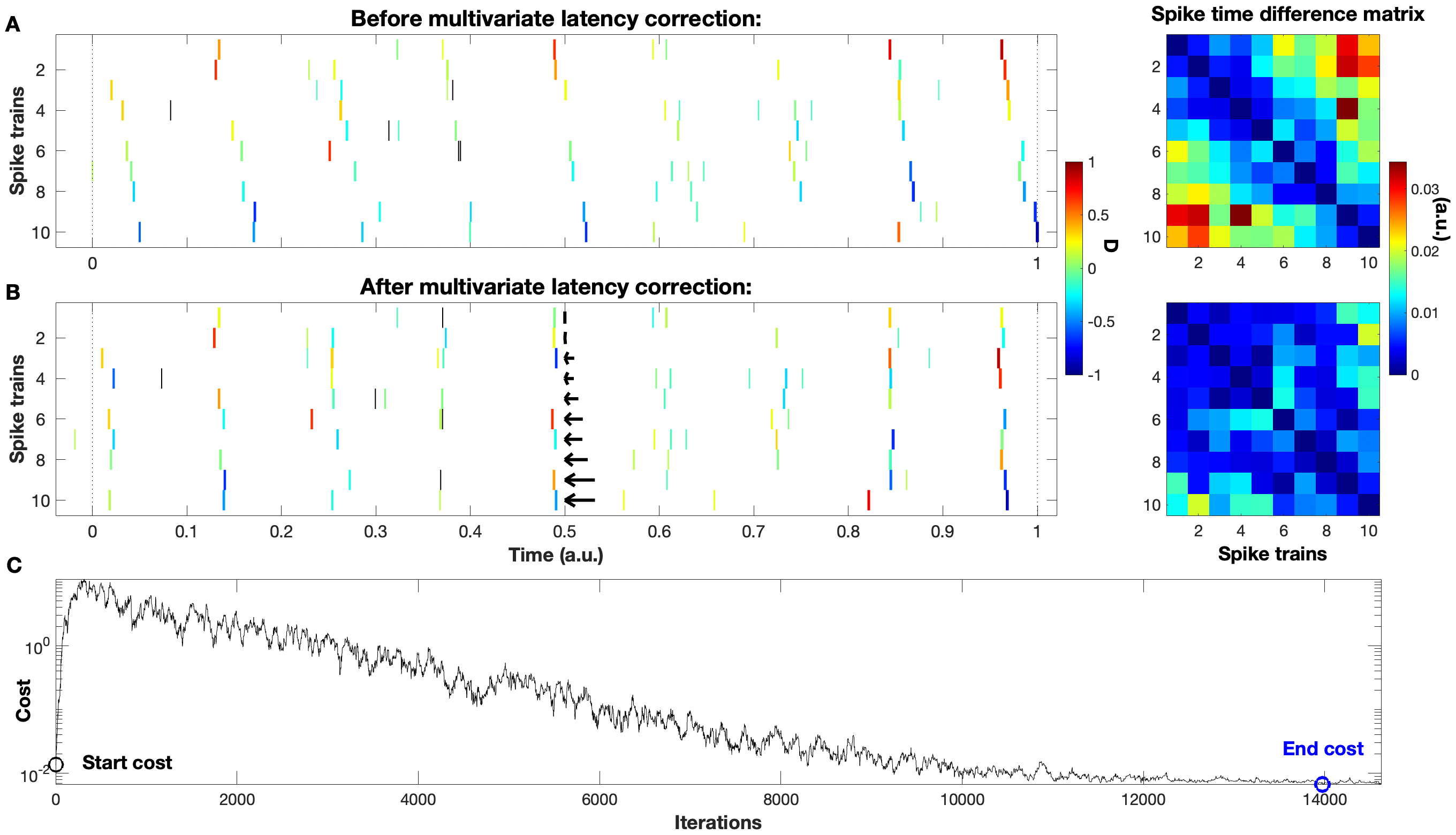}
	\caption{Similar to Fig. \ref{F1_Synfire_chain} but this time we show a more realistic dataset superimposed with some unreliability, jitter, and background noise. Clearly the latency corrected spike trains in rasterplot B exhibit a much larger degree of synchrony than the ones in rasterplot A. Notice that all three of the aforementioned sources of noise are still present after the correction. On the right, the values of the spike time difference matrix decrease considerably. In subplot C we also display the cost function over the course of the simulated annealing. It consists of the usual two parts, a short initial increase from the start cost (marked by a black circle) and a decrease that slowly convergences towards the end cost (defined as the minimum value of the cost function and here marked by a blue circle).}
\label{F2_Simple_Example_Full_Plot}
\end{figure*}

However, datasets in real life are not as clean and we encounter `disturbances' such as incomplete global events, jitter, and background noise. Such a more realistic example is shown in Fig. \ref{F2_Simple_Example_Full_Plot}. Going further, datasets can contain different duration of global events and different intervals between subsequent spikes (non-monotonous propagation), until in the end we arrive at spike trains sets without any clear propagation structure. For the perfect synfire chain of Fig. \ref{F1_Synfire_chain} it is enough to consider $N-1$ entries of the matrix and ignore the others, but under more realistic conditions the solution obtained this way becomes suboptimal and a more general and sophisticated approach is needed.
%
%

For this we propose simulated annealing \cite{Dowsland12}, an heuristic approach that uses an iterative directed random walk to find the optimum (here minimum) of a cost function within the vast search space. 

Starting from the initial cost value before the latency correction (\textit{start cost}), all iterations which decrease the cost function are accepted while the likelihood of accepting iterations which increase the cost function is getting lower and lower according to a slow cooling scheme which ensures a certain degree of convergence. However, since this probability remains always positive, simulated annealing, in contrast to a steepest descent (or `greedy') algorithm, has the ability to recover from local (but non-global) minima. Iterations last until the cost no longer changes or until a predefined end temperature is reached. As \textit{end cost} after the latency correction we use the minimum cost obtained over the course of the simulated annealing.

In multivariate latency correction the search space is composed of all possible shifts of the spike trains relative to each other and in principle this space is infinitely large. The cost function to be minimized is the average value of the upper right tridiagonal part of the spike time difference matrix (which thus takes into account all elements of the matrix):
\begin{equation} \label{eq:Cost-function}
	c = \langle \Delta_< \rangle.
\end{equation}
Within each iteration a randomly selected spike train is shifted by a randomly selected time interval. As a special trick, to facilitate convergence the shift values are drawn from a Gaussian distribution that gets narrower the closer we get to the optimal solution: at each iteration its standard deviation is set to the current cost value. The typical course of the cost function during the simulated annealing is shown for the more realistic example of Fig. \ref{F2_Simple_Example_Full_Plot}.

%
\subsection{SPIKE-order and the Synfire Indicator} \label{ss:Methods-SPIKE-order-Synfire-Indicator}

In this Section we introduce the concepts and the quantities needed to later define a quantitative criterion for the suitability of datasets for our multivariate latency correction algorithm.

SPIKE-Synchronization is invariant to which of the two spikes within a coincidence pair is leading and which is following. To take the temporal order of the spikes into account we developed the SPIKE-Order approach \cite{Kreuz17} which allows to sort the spike trains from leader to follower and to evaluate the consistency of the preferred order via the Synfire Indicator.

Following  \cite{Kreuz17}, we first define the bivariate anti-symmetric SPIKE-Order indicator 
\begin{eqnarray} \label{eq:SPIKE-Order-Spike}
	D_i^{(n,m)} & = & C_i^{(n,m)} \cdot \sign (t_{j'}^{(m)} - t_i^{(n)}) \nonumber \\
	D_{j'}^{(m,n)} & = & C_{j'}^{(m,n)} \cdot \sign (t_i^{(n)} - t_{j'}^{(m)}) = - D_i^{(n,m)},
\end{eqnarray}
which assigns to each spike $i$ either a $1$ or a $-1$ depending on whether the respective spike is leading or following the coincident spike $j'$ in the other spike train (cf. Eq. \ref{eq:Matching-Spike}).

SPIKE-Order distinguishes leading and following spikes, and is thus used to colorcode the individual spikes on a leader-to-follower scale (see, e.g., the rasterplots in Fig. \ref{F1_Synfire_chain}). It can also be employed to sort the \textit{spike trains} by means of the cumulative and anti-symmetric SPIKE-Order matrix
\begin{equation} \label{eq:SPIKE-Order-Matrix}
    D^{(n,m)} = \sum_i D_i^{(n,m)}
\end{equation}
which quantifies the temporal relationship between spike trains $n$ and $m$. 
If $D^{(n,m)}>0$ spike train $n$ is leading $m$, while $D^{(n,m)}<0$ means $m$ is the leading spike train.
For a spike train order in line with the synfire property (i.e., exhibiting consistent repetitions of the same global propagation pattern), we thus expect $D^{(n,m)} > 0$ for $n<m$. Therefore, the overall SPIKE-Order can be constructed as
\begin{equation} \label{eq:SPIKE-Order}
 	D_< = \sum_{n<m} D^{(n,m)},
\end{equation}
i.e.\ the sum over the upper right tridiagonal part of the matrix $D^{(n,m)}$.
%
%
\begin{figure*}[!ht]
\includegraphics[width=\linewidth]{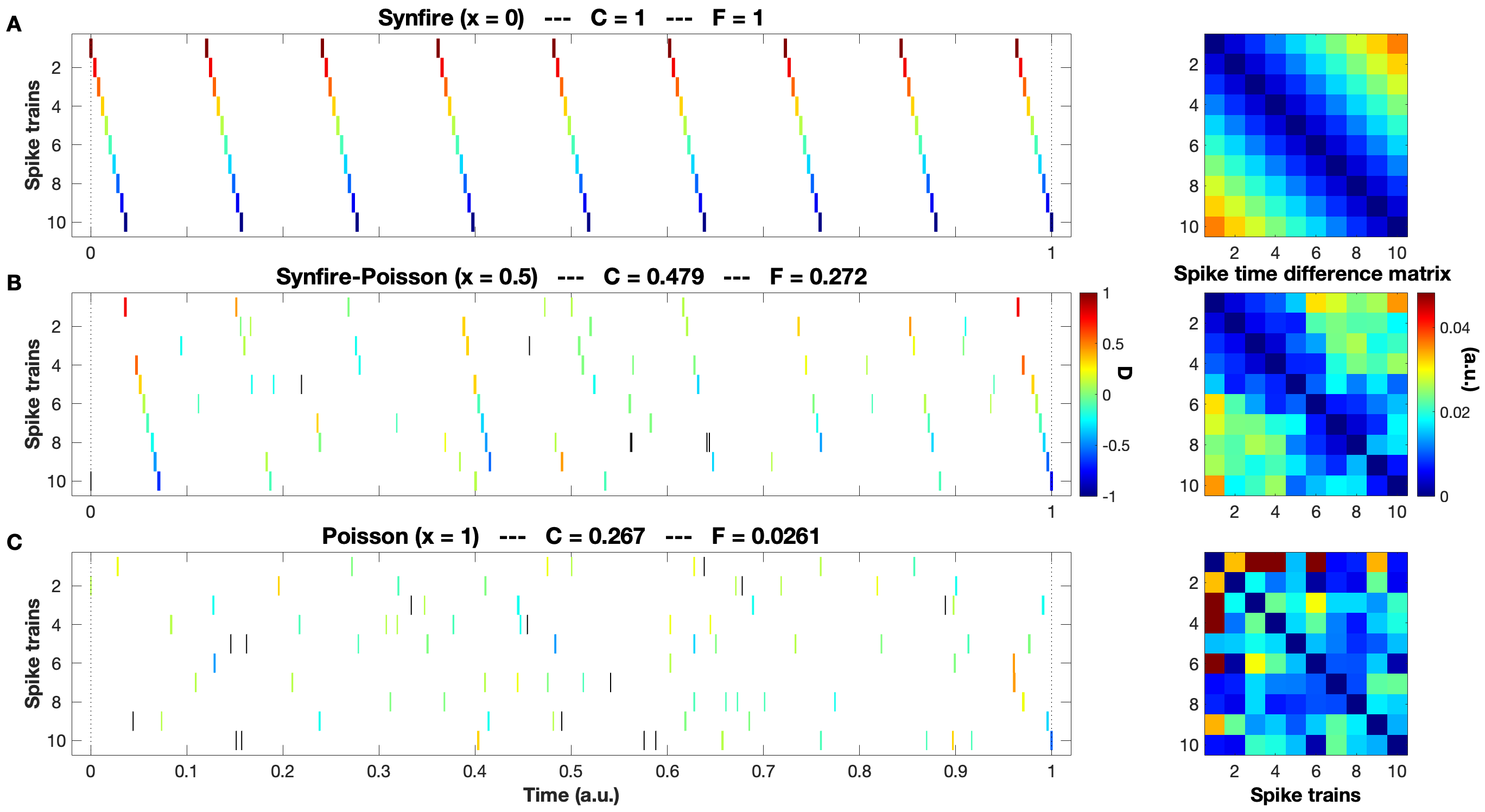}
\caption{Simulated data with varying mixing parameter $x$: From a perfect synfire chain (subplot A, $x = 0$) via the halfway case (subplot B, $x = 0.5$) to a pure Poisson process (subplot C, $x = 1$). For each example, we state the values of the SPIKE-synchronization $C$ and the Synfire Indicator $F$. Note how the spike time difference matrices get more and more irregular. While the trend underlying the perfect regularity from A (monotonous increase as one moves away from the diagonal) is still perceptible in B, in C there is no apparent order anymore.}\label{fig:F3_ResSim1_Mixing-3Examples}
\end{figure*}

Finally, normalizing by the total number of possible coincidences yields the Synfire Indicator:
\begin{equation} \label{eq:Synfire-Indicator-F}
	F = \frac{2 D_<}{(N-1) M}. 
\end{equation}
This measure quantifies to what degree coinciding spike pairs with correct order prevail over coinciding spike pairs with incorrect order, or, in other words, to what extent the spike trains in their current order resemble a consistent synfire pattern.
Accordingly, maximizing the Synfire Indicator $F_\varphi$ as a function of the spike train order $\varphi(n)$ finds the sorting of the spike trains from leader to follower such that the sorted set $\varphi_s$ comes as close as possible to a perfect synfire pattern:
\begin{equation} \label{eq:Sorted-Order}
	\varphi_s: F_{\varphi_s} = \max_\varphi \{F_\varphi\} = F_s.
\end{equation}

Whereas the Synfire Indicator $F_\varphi$ for any spike train order $\varphi$ is normalized between $-1$ and $1$, the optimized Synfire Indicator $F_s$ can only attain values between $0$ and $1$. But from Eq. \ref{eq:SPIKE-Order-Spike} if follows that since the order is only evaluated among those spikes that match, the actual upper bound for any given dataset is the value of SPIKE-synchronization $C$ (Eq. \ref{eq:SPIKE-synchronization}). A perfect synfire pattern results in $F_s=1$, while sufficiently long Poisson spike trains without any synfire structure yield $F_s \gtrsim 0$. 

In this article, for both the simulated and the real datasets before the multivariate latency correction we first sort the spike trains from leader to follower. The optimization procedure that we use to find the best spike train order is again based on simulated annealing (details can be found in \cite{Kreuz17}). While this sorting is not necessary for the correction itself, it renders the resulting rasterplots more intuitive and easier to read. For simplicity we refer to the Synfire Indicator of the sorted spike trains as $F$.

As we will show in Section \ref{ss:Results-SimExp}, the Synfire Indicator $F$ serves as main criterion for the suitability of our algorithm that can be evaluated for a given dataset before the multivariate latency correction is actually applied. On the other hand, once we have performed the correction we would also like to quantify how successful it actually has been. As measure of its performance we use the relative cost improvement (in percent) defined as
\begin{equation} \label{eq:Relative_cost_improvement}
	I = \frac{c_{start}-c_{end}}{c_{start}}*100,
\end{equation}
the normalized change in cost (i.e. the mean spike time difference, see Eq. \ref{eq:Cost-function}) between before ($c_{start}$) and after ($c_{end}$) the correction. For comparison purposes we also define the shift cost $c_{shift}$ as the cost that is obtained after shifting the spike trains according to the delays in the first row of the spike time difference matrix (without performing simulated annealing and while ignoring all other entries of that matrix). As we have seen in Fig. \ref{F1_Synfire_chain}, for a perfect synfire chain this value is zero.

%
\section{Results} \label{s:Results}

First, in Section \ref{ss:Results-Sim} we investigate the performance of our new method for multivariate latency correction in a controlled setting using simulated data that cover the whole range from a perfect synfire chain to pure Poisson spike trains. Then we apply the algorithm to neurophysiological datasets, cortical propagation patterns recorded via wide-field calcium imaging from mice before and after stroke (Section \ref{ss:Results-Exp}). Finally, in Section \ref{ss:Results-SimExp} we perform simulations of these experimental data which allow us to extend the parameter range and derive a criterion which determines whether a given dataset is a suitable candidate for our algorithm.

\subsection{Simulated data} \label{ss:Results-Sim}

Before applying our method to the experimental datasets described in Section \ref{s:Data} we test it on controlled data with known ground truth. To this aim, we introduce a mixing parameter $x$, that is used to interpolate between the two extremes of perfect synfire chain ($x = 0$) and pure Poisson spike trains ($x = 1$). This mixing parameter is increased from $0$ to $1$ in steps of $0.05$ and for every one of these $21$ values we generate $N = 10$ spike trains. For each of these spike trains we select a fraction $x$ of spikes from a perfect synfire chain (with $M_n$ = 9 spikes) and a fraction $1-x$ from a pure Poisson spike train (with an expectation value of $\langle M_n \rangle = 9$ spikes). This way each spike train becomes a superposition of a synfire chain and a Poisson spike train with the relative contribution determined by the mixing parameter. In Fig. \ref{fig:F3_ResSim1_Mixing-3Examples} we show the two extremes and in between the halfway case. For each example we also report the values of the SPIKE-synchronization $C$ (cf. Section \ref{ss:Methods-Spike-Matching-SimAnn}) and the Synfire Indicator $F$ (Section \ref{ss:Methods-SPIKE-order-Synfire-Indicator}).
%
%
\begin{figure}[!ht]
\begin{center}
\includegraphics[width=\linewidth]{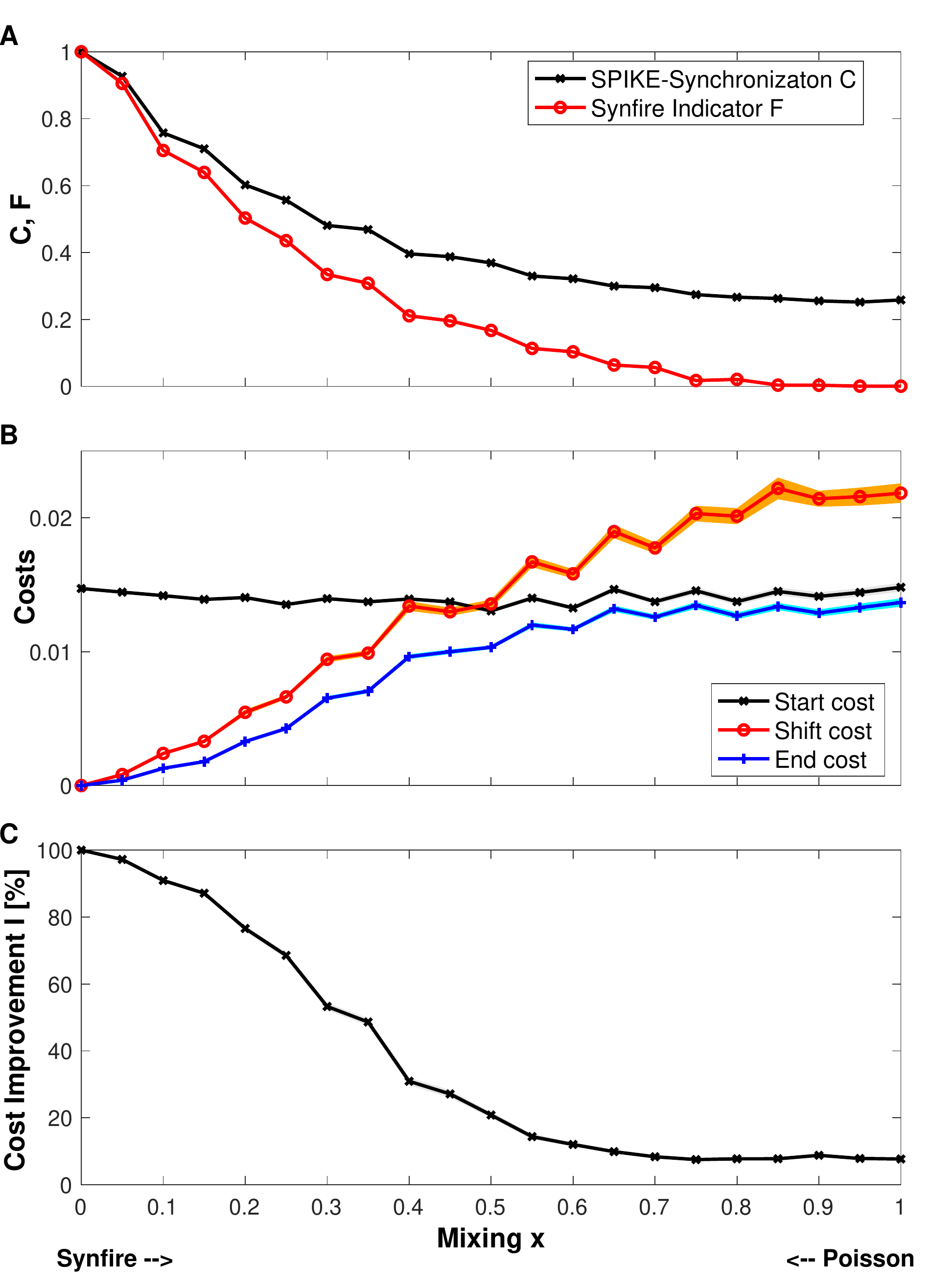}
\caption{Simulated data: SPIKE-Synchronization $C$ and Synfire Indicator $F$ (subplot A), start, shift and end costs (subplot B), as well as relative cost improvement in percent (subplot C) versus the mixing parameter $x$.
Shaded areas indicate the standard error of the mean (which in some cases is hardly visible). In A both SPIKE-Synchronization and the Synfire Indicator decrease rather monotonously with $x$. At the same time, $C$ acts as upper bound for $F$. In B around a mixing value of $x = 0.5$ the end cost is not that much lower than the start cost anymore, and accordingly in C the cost improvement starts to level off towards a rather low value below $10 \%$. All graphs show averages over 100 independent realizations for each mixing value. While the main synfire chain always remains the same (like the one shown in Fig. \ref{fig:F3_ResSim1_Mixing-3Examples}A), for $x > 0$ the stochastic parts vary for each realization: different synfire spikes are omitted and different Poisson spikes are added.}
\label{fig:F4_ResSim1_Dependence_on_mixing}
\end{center}
\end{figure}	

Fig. \ref{fig:F4_ResSim1_Dependence_on_mixing}A reports $C$ and $F$ in dependence of the mixing parameter. Both values decrease with increasing mixing parameter. For the perfect synfire chain ($x = 0$) they start with their maximum value of one, whereas for pure Poisson spike trains ($x = 1$) they reach their lowest value which due to remaining random coincidences and persistent random order, respectively, gets quite close to but does not reach zero. As already mentioned in Section \ref{ss:Methods-SPIKE-order-Synfire-Indicator}, by definition SPIKE-synchronization $C$ is an upper limit for the Synfire Indicator $F$.

In Fig. \ref{fig:F4_ResSim1_Dependence_on_mixing}B we display the behaviour of the three relevant cost values. The start cost hovers around some intermediate value and seems to be quite independent of $x$. In the case of a perfect synfire chain ($x = 0$) the end cost is actually the shift cost since we refrain from running the simulated annealing because we have already reached the optimum value of zero. However, for all positive values of the mixing parameter the end cost obtained via simulated annealing outperforms the shift cost, as a tendency the more so the higher $x$. This shows that while the shift cost is very fast to calculate, in general it gives only suboptimal results. Roughly starting from the halfway point ($x = 0.5$) it is actually even worse than the start cost.

The end cost also consistently improves on the start cost and this is quantified more directly in Fig. \ref{fig:F4_ResSim1_Dependence_on_mixing}C where we show the relative cost improvement $I$ (Eq. \ref{eq:Relative_cost_improvement}) in dependence of the mixing parameter. The improvement is highest for the synfire chain where the correction yields the perfect result and then it slowly drops off until somewhere between $x = 0.5$ and $x = 0.7$ it reaches a plateau where the improvement is still positive but rather low. This cutoff range of the mixing parameter corresponds to SPIKE-Synchronization values of $C = 0.28 ... 0.35$ and Synfire Indicator values of $F = 0.04 ... 0.14$ (see Fig. \ref{fig:F4_ResSim1_Dependence_on_mixing}A). 
%
%
\begin{figure*}[!ht] 
\includegraphics[width=\linewidth]{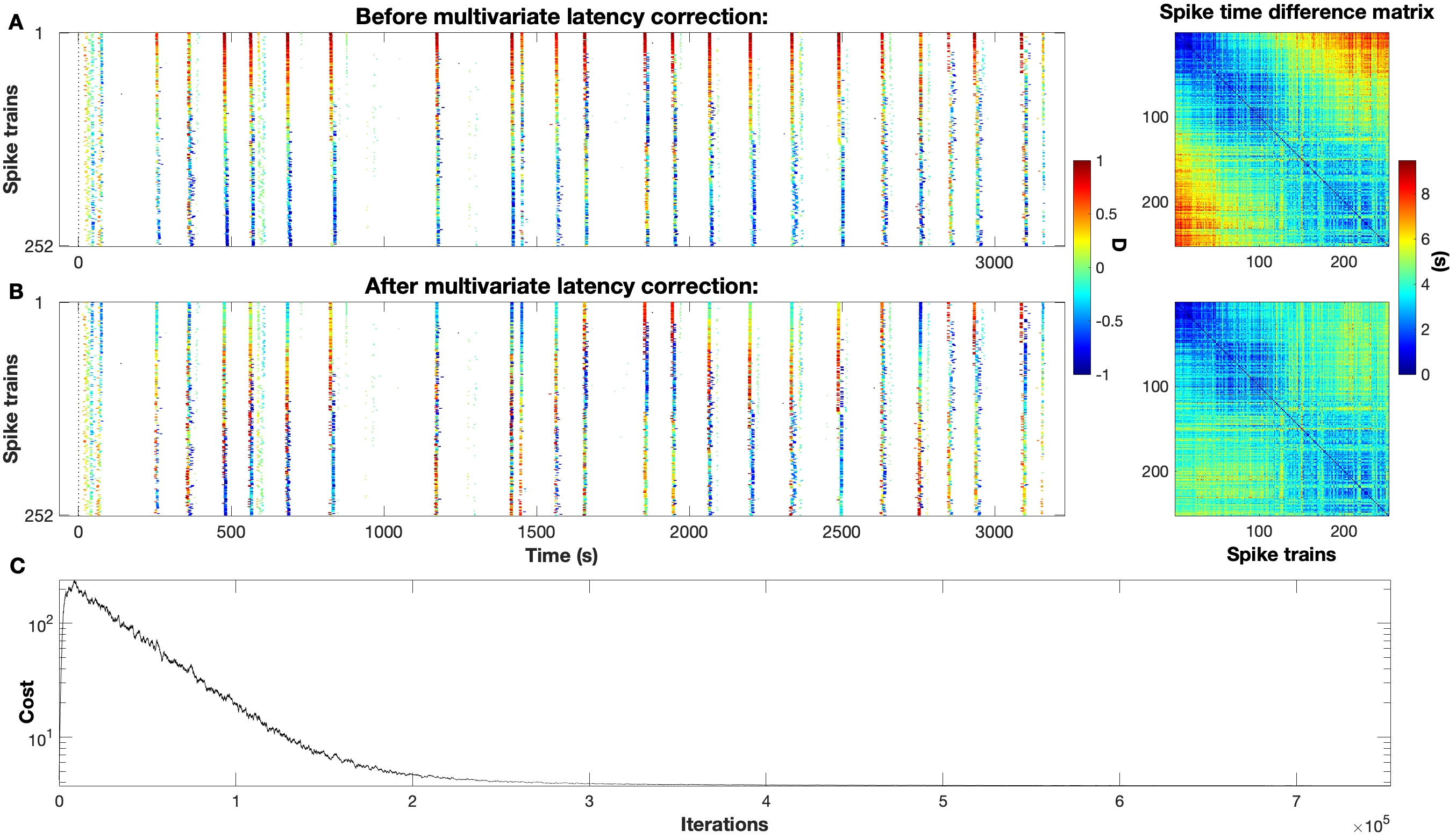}
\caption{Example of latency correction in an experimental dataset: The spikes in the rasterplot are the upwards threshold crossings of the $252$ calcium traces recorded in vivo during motor training from a healthy control mouse (mouse $29$, day $9$). The plot follows the structure from Fig. \ref{F2_Simple_Example_Full_Plot}. The order in different global events varies but typically there is a rather high level of consistency as can be seen by the rainbow-like color patterns of the events that mostly go from red (leading spikes) to blue (following spikes). On this dataset (SPIKE-synchronization $C = 0.867$, Synfire Indicator $F = 0.366$) the effect of the latency correction can be seen most clearly in the outer corners of the spike time difference matrix, the intervals between the first leaders and the last followers get reduced considerably. Overall, the resulting cost improvement is $I = 10.98 \%$.}
\label{fig:F5_ResExp_Data_Example1_099_M29_D9}
\end{figure*}

%
\subsection{Experimental data} \label{ss:Results-Exp}

After validating our algorithm on controlled data with known ground truth, we now show its effectiveness in a real life application to neurophysiological datasets. For this we choose spatiotemporal propagation activity in the cortex of mice observed with in vivo calcium imaging before and after the induction of a stroke \cite{mascaro2019combined, Cecchini21} (see Section \ref{s:Data}).
%
%
\begin{figure*}[!ht]
\includegraphics[width=\linewidth]{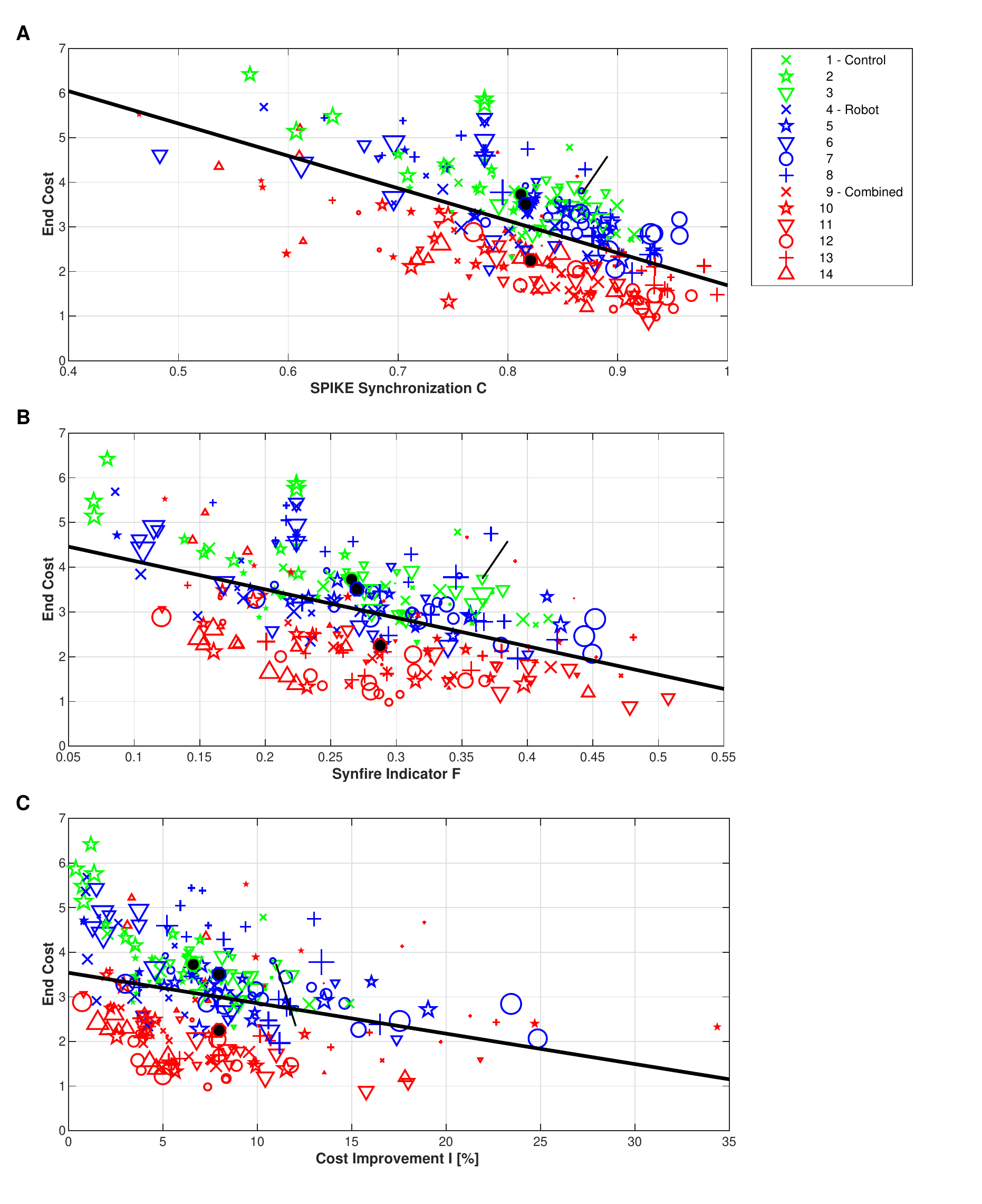}
\vspace{-10mm}
\caption{Statistics for experimental data: End cost versus SPIKE-Synchronization (subplot A), the Synfire Indicator (subplot B) and the relative cost improvement (subplot C) for all $260$ datasets. Colors indicate groups, symbols distinguish mice, and the larger the marker, the later the day of the recording. For each group we also added the center of mass indicated by a larger marker with a black center. The thick black lines represent a linear fit for all datasets together, independent of the group. The short black lines indicate the values for the example dataset shown in Fig. \ref{fig:F5_ResExp_Data_Example1_099_M29_D9}. The end cost is anticorrelated with both SPIKE-Synchronization and the Synfire Indicator as well as with the relative cost improvement.}
\label{fig:F6_ResExp_RealData_Stats1}
\end{figure*}
%
%
\begin{figure*}[!ht]
\includegraphics[width=\linewidth]{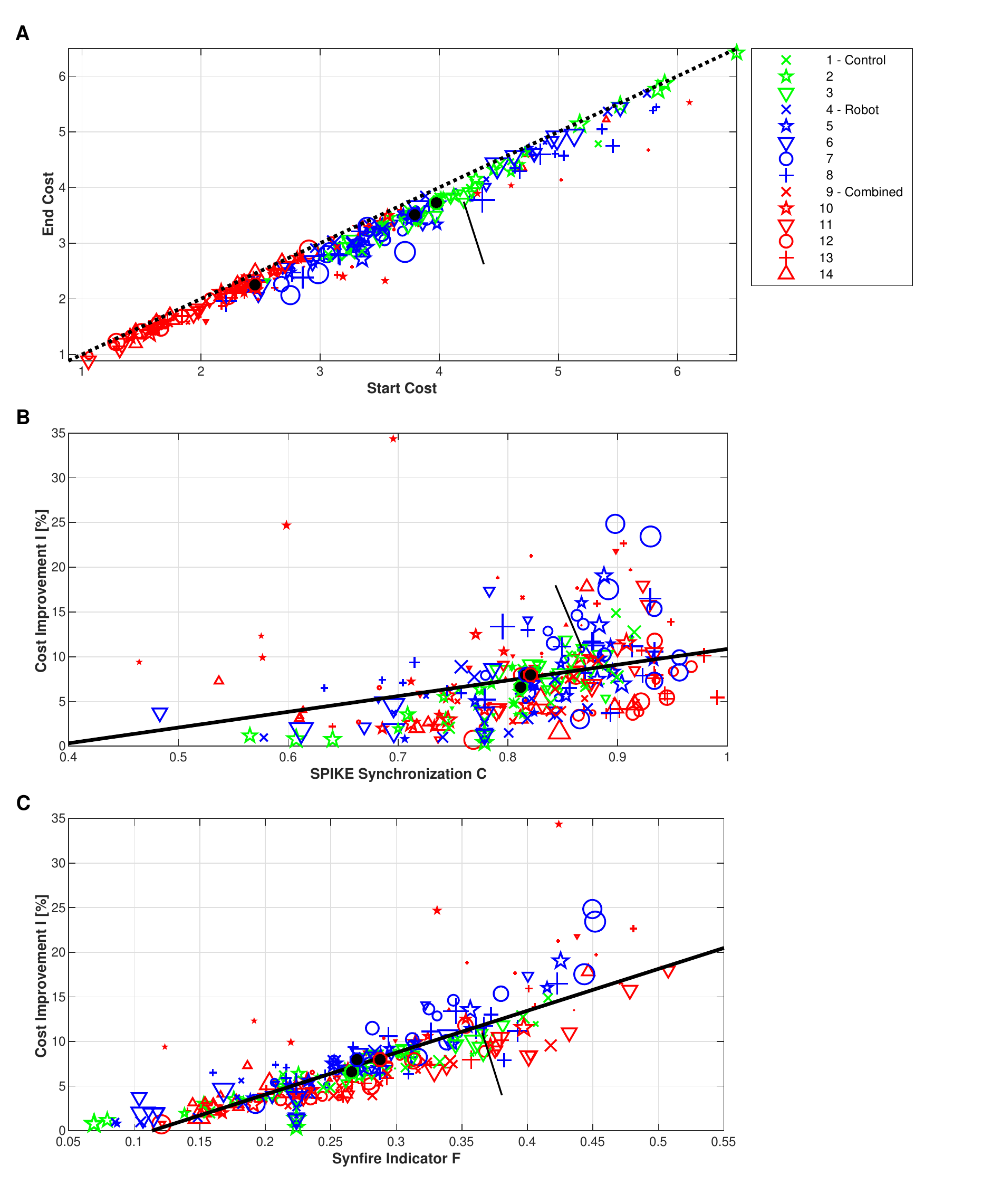}
\vspace{-10mm}
\caption{Performance of algorithm on experimental data: End cost versus start cost (subplot A) as well as relative cost improvement versus SPIKE-Synchronization (subplot B) and versus the Synfire Indicator (subplot C) for all $260$ datasets. Layout as in Fig. \ref{fig:F6_ResExp_RealData_Stats1}. In A the diagonal (corresponding to unchanged costs) is marked by a dashed black line. The effect of the latency correction on the costs is most pronounced in the middle range. The relative cost improvement is much less correlated with SPIKE-Synchronization (B) than with the Synfire Indicator (C).}
\label{fig:F7_ResExp_RealData_Stats2}
\end{figure*}

In our previous work \cite{Cecchini21} we have analyzed these datasets and applied three novel indicators (angle, duration and smoothness) based on asymmetric measures of directionality to the observed global activation patterns in order to track damage and functional recovery during various rehabilitation paradigms. Here we would like to follow a complementary approach and look at the similarity of the activity during the course of the propagation. The aim is to investigate whether this new type of analysis can help us to distinguish the three different groups of mice, Control, Robot and Combined. But first we have to correct for the systematic latency caused by the finite propagation speed.

Fig. \ref{fig:F5_ResExp_Data_Example1_099_M29_D9}A shows a typical dataset (this one from a healthy control mouse) with a complete set of $12 \times 21 = 252$ spike trains. This particular rasterplot exhibits about $25$ rather complete global events plus a few incomplete events and a limited amount of background noise. Since spike trains are already sorted, the global events mostly observe the rainbow pattern from red (leading spikes) to blue (following spikes). Typically, the matched spikes from the very first and the very last spike trains are furthest apart and, accordingly, the highest values in the spike time difference matrix are found in the corners away from the diagonal. On the other hand, the data are quite far from a perfect synfire chain. Apart from the incompleteness of some of the global events and the background noise there is also quite a bit of variability in the order within the events. In fact, because of this the value of SPIKE-synchronization for this example is $C = 0.867$ while the Synfire Indicator is only $F = 0.366$, both clearly below their maximum value of $1$.

After we run our multivariate latency correction it is in particular the distances between the first leaders and the last followers (the matrix elements furthest away from the diagonal) that are greatly reduced (Fig. \ref{fig:F5_ResExp_Data_Example1_099_M29_D9}B). The overall cost value falls from $4.21$ to $3.74$, a drop that corresponds to a relative cost improvement of $11 \%$. It takes slightly more than $750.000$ iterations to reach this improvement (Fig. \ref{fig:F5_ResExp_Data_Example1_099_M29_D9}C).

Next, we look at the statistics over all $260$ datasets. In Figs. \ref{fig:F6_ResExp_RealData_Stats1}A and \ref{fig:F6_ResExp_RealData_Stats1}B we plot the end cost $c_{end}$ (i.e. the average value of the spike time difference matrix after the multivariate latency correction) versus SPIKE-Synchronization $C$ and the Synfire Indicator $F$, respectively. Comparing the different groups, the end cost is lowest (similarity is highest) for the Combined group followed by the Robot and the Control group, but there is quite some overlap between the different distributions. For all three groups separately, and combined, there is a tendency that the cost is lower for higher values of both SPIKE-Synchronization and the Synfire Indicator (overall, the least square fits show linear correlations of $R = -0.594$ for $C$ and $R = -0.501$ for $F$).

Fig. \ref{fig:F6_ResExp_RealData_Stats1}C marks the transition from the description of the data to the characterization of the performance of our algorithm. \tcb{The end cost quantifies the corrected similarity of the datasets and is thus the main value that we are interested in from a data point of view. On the other hand, the relative cost improvement characterizes the relative effect of the method in correcting the latency for a given set of data.} \tcr{Fig. \ref{fig:F6_ResExp_RealData_Stats1}C relates these two quantities and indicates that they are slightly anticorrelated with a correlation coefficient of $R = - 0.307$. This anticorrelation can be expected from the definition of the relative cost improvement in Eq. \ref{eq:Relative_cost_improvement} where the end cost enters as a subtrahend. The fact that the absolute correlation value is not higher is due to the influence of the start cost which then determines the relative change.}

In order to analyze the performance of the algorithm in more detail, in Fig. \ref{fig:F7_ResExp_RealData_Stats2}A we compare the cost after the latency correction versus the cost before the latency correction, again for all $260$ datasets. For a correction that would just result in a percentual offset, all values would lie on a diagonal parallel to the main diagonal. As can be seen from the many off-diagonal values and the occasional larger outlier this is clearly not the case. We also find that the points being furthest away from the diagonal are concentrated in an intermediate range suggesting that the algorithm works best in cases with intermediate synchrony.

Figs. \ref{fig:F7_ResExp_RealData_Stats2}B and \ref{fig:F7_ResExp_RealData_Stats2}C display the relative cost improvement in dependence of SPIKE-Synchronization and the Synfire Indicator, respectively. Here it is much more difficult to separate the three different groups than in Fig. \ref{fig:F6_ResExp_RealData_Stats1} which demonstrates that the algorithm performs for all three groups equally well. 

When we look at the dependence of the relative cost improvement on the SPIKE-synchronization $C$ we find a rather modest linear correlation of $R = 0.319$. However, the most pronounced linear correlation over all the datasets is obtained for the Synfire Indicator: the larger $F$ the bigger the relative cost improvement and here we obtain an astonishing $R = 0.822$. Thus, the primary influence on the performance of the algorithm is the Synfire Indicator, while the role of SPIKE-synchronization merits further investigation.


%
%
\subsection{Simulations of experimental data} \label{ss:Results-SimExp}

As a last step, we simulate the experimental data analyzed in the previous Section \ref{ss:Results-Exp}. With this we have two major objectives in mind: (i) to extend the parameter range covered by the experimental data and (ii) to control the simulations such that we can isolate the influence of the two most important characterizing quantities SPIKE-synchronization and Synfire Indicator in a more systematic way, something which can not be done with the random and arbitrary distributions of the real data.
%
%
\textit{\begin{figure*}[!thb]
    \centering
    \begin{subfigure}[b]{0.485\textwidth}
        \centering
        \includegraphics[width=\textwidth]{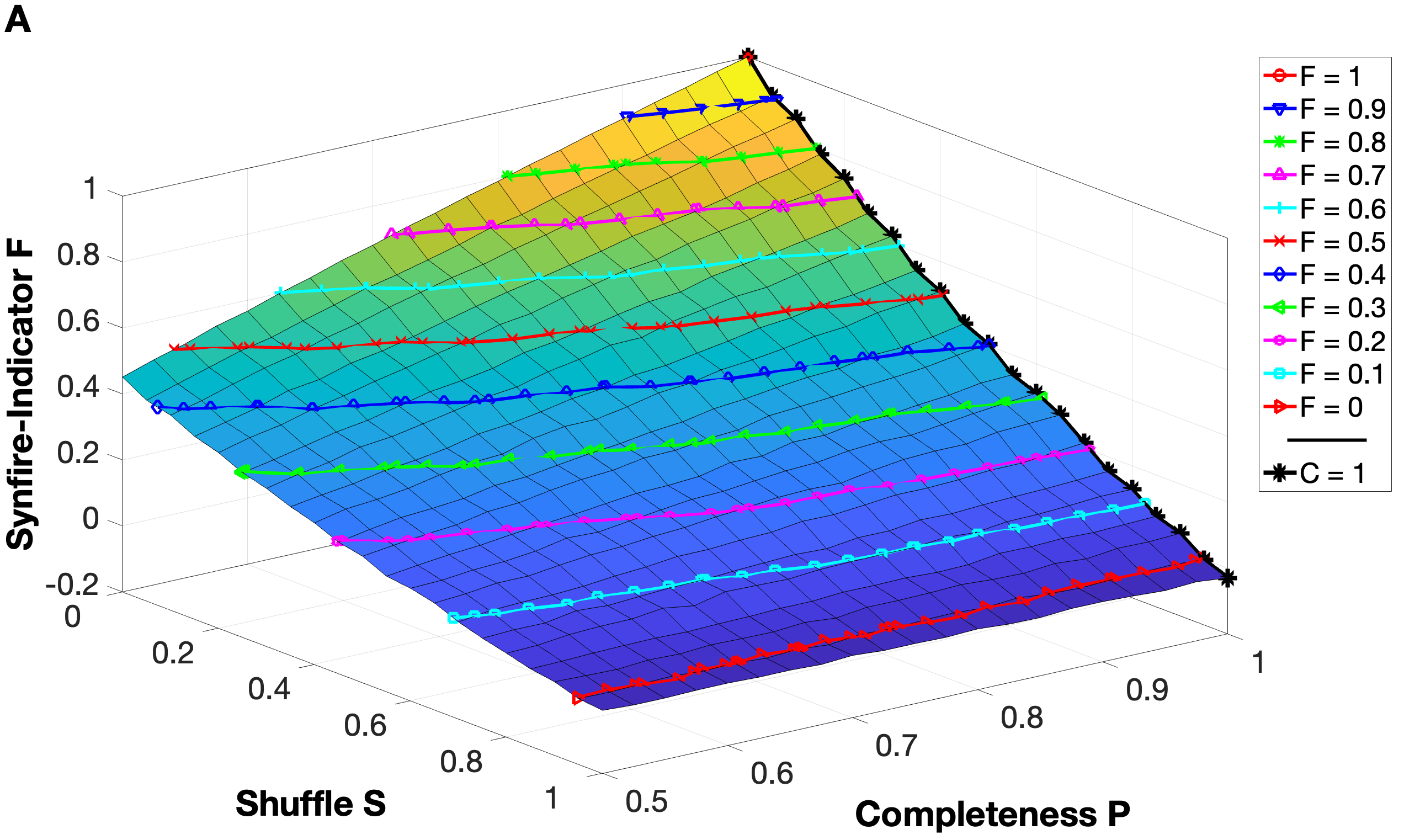}
    \end{subfigure}
    \hfill
    \begin{subfigure}[b]{0.485\textwidth}  
        \centering 
        \includegraphics[width=\textwidth]{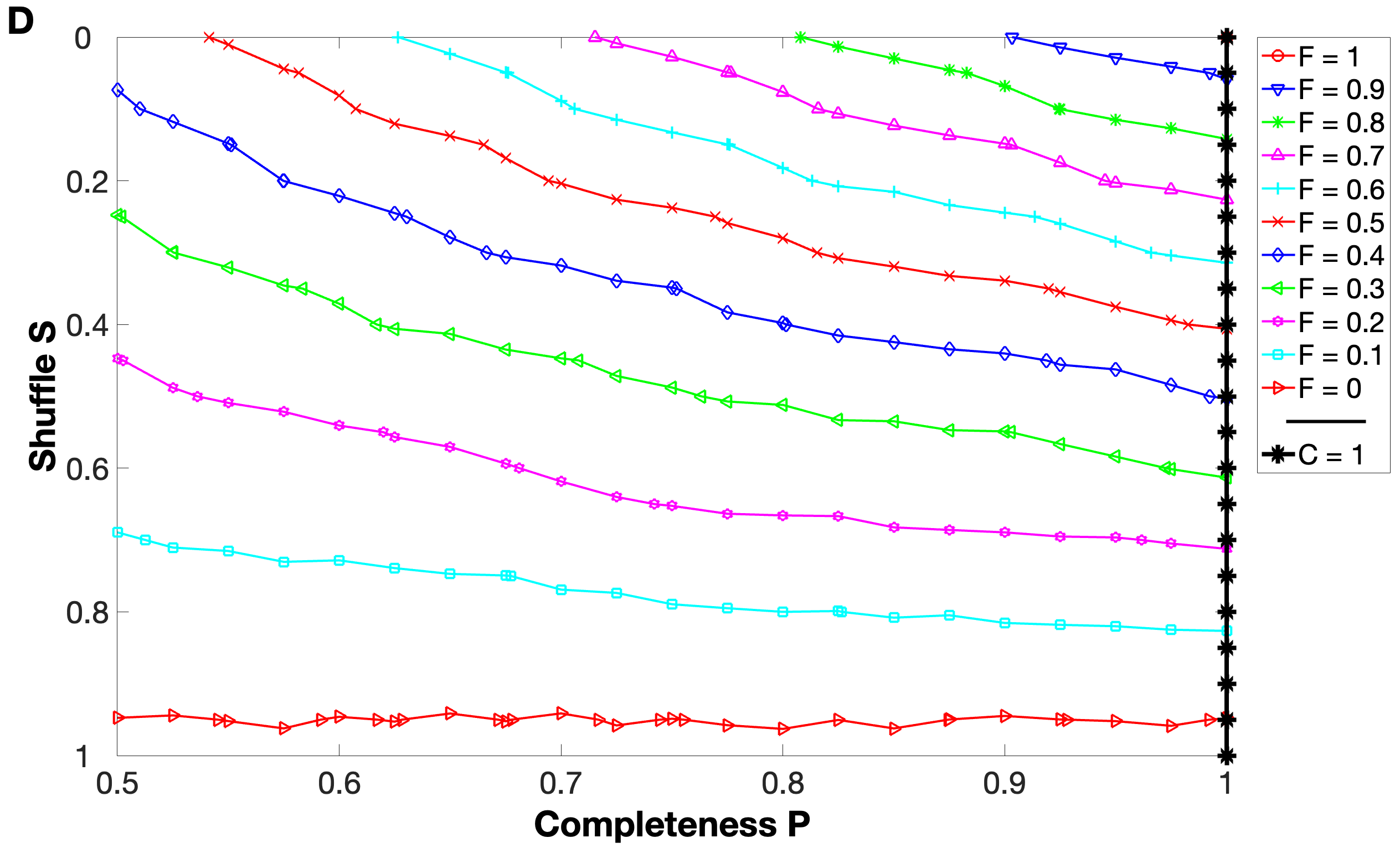}
    \end{subfigure}
    \vspace{-2mm}
    \vskip \baselineskip
    \begin{subfigure}[b]{0.485\textwidth}   
        \centering 
        \includegraphics[width=\textwidth]{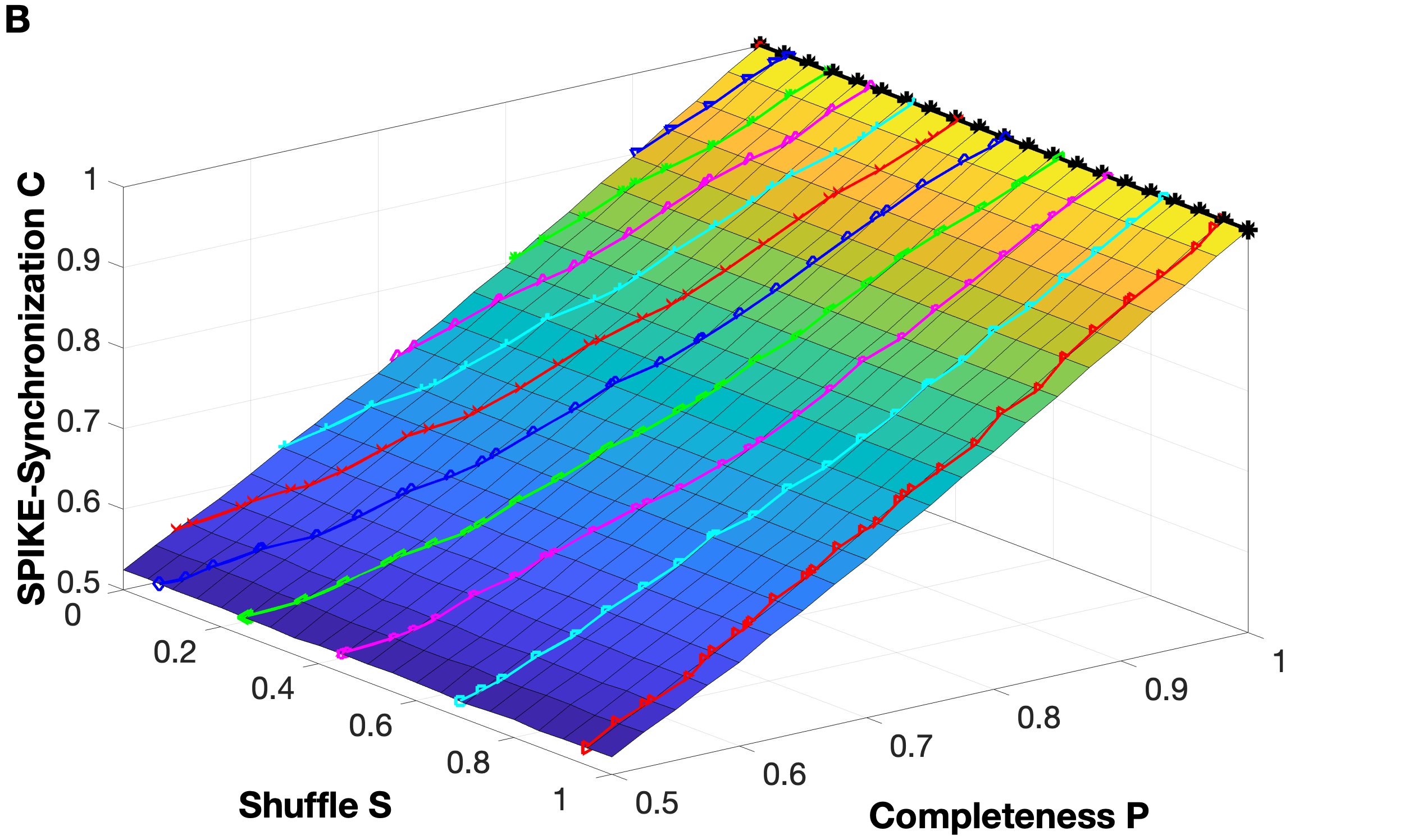}
    \end{subfigure}
    \hfill
    \begin{subfigure}[b]{0.485\textwidth}   
        \centering 
        \includegraphics[width=\textwidth]{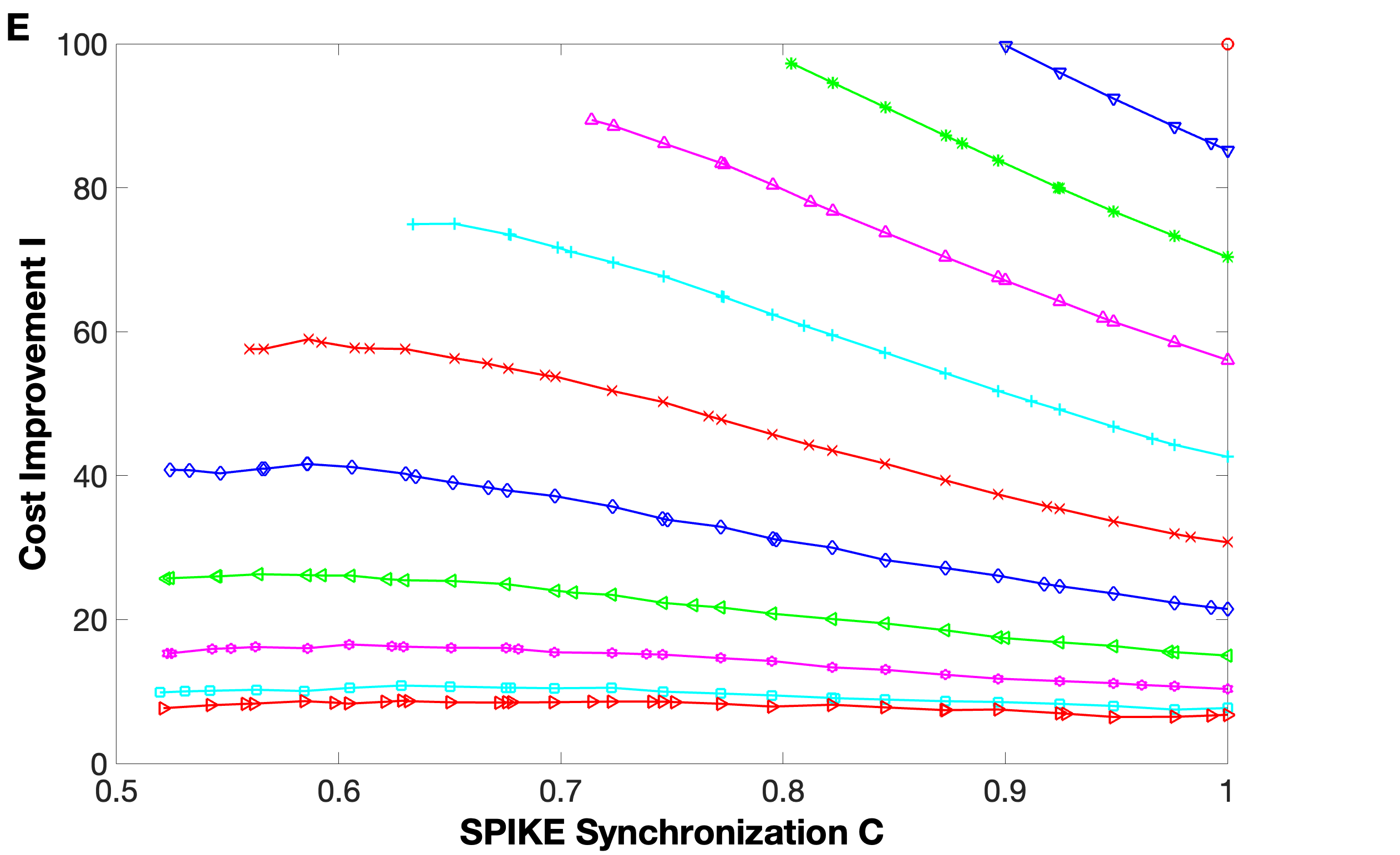}
    \end{subfigure}
    \vspace{-2mm}
    \vskip\baselineskip
    \begin{subfigure}[b]{0.485\textwidth}   
        \centering 
        \includegraphics[width=\textwidth]{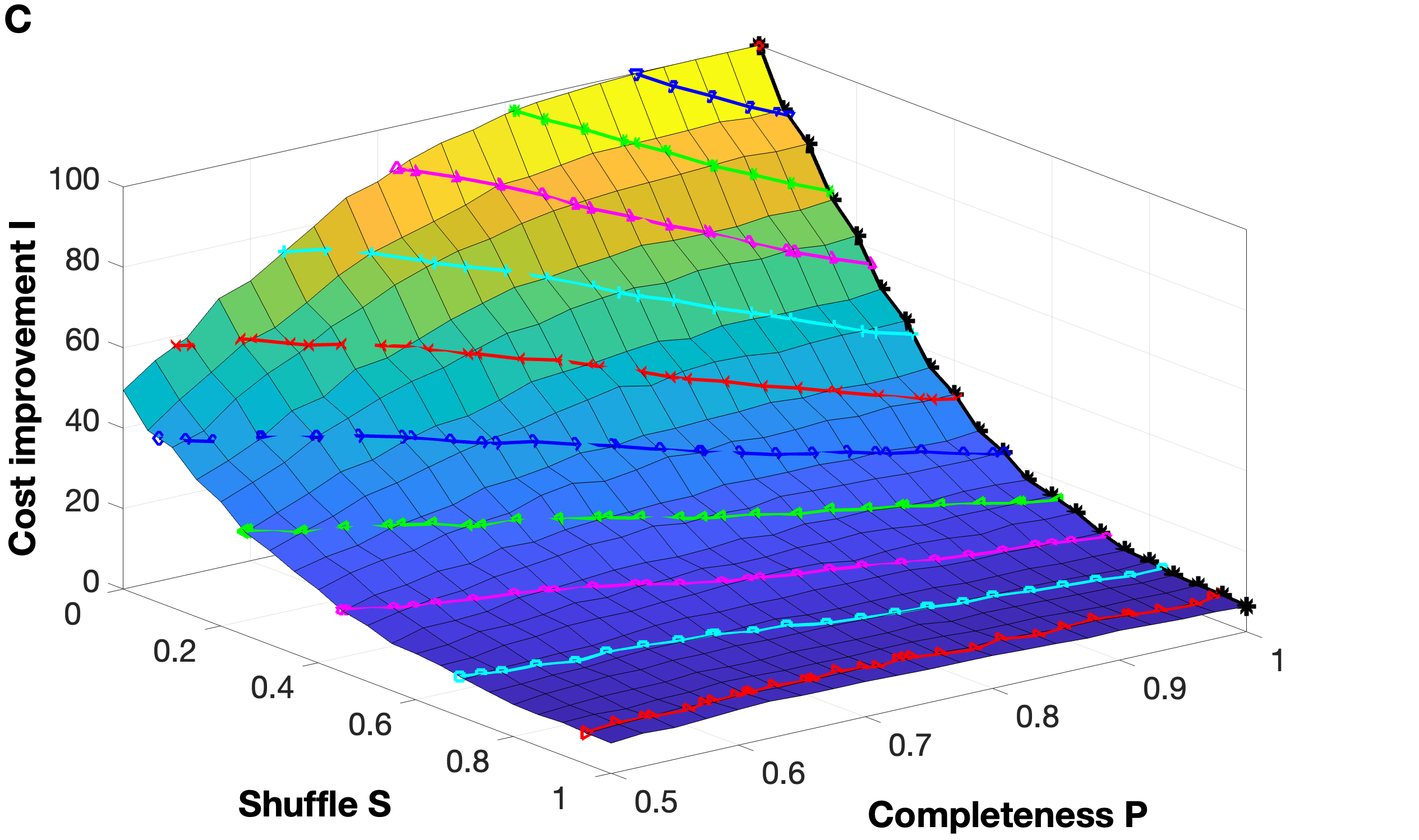}
    \end{subfigure}
    \hfill
    \begin{subfigure}[b]{0.485\textwidth}   
        \centering 
        \includegraphics[width=\textwidth]{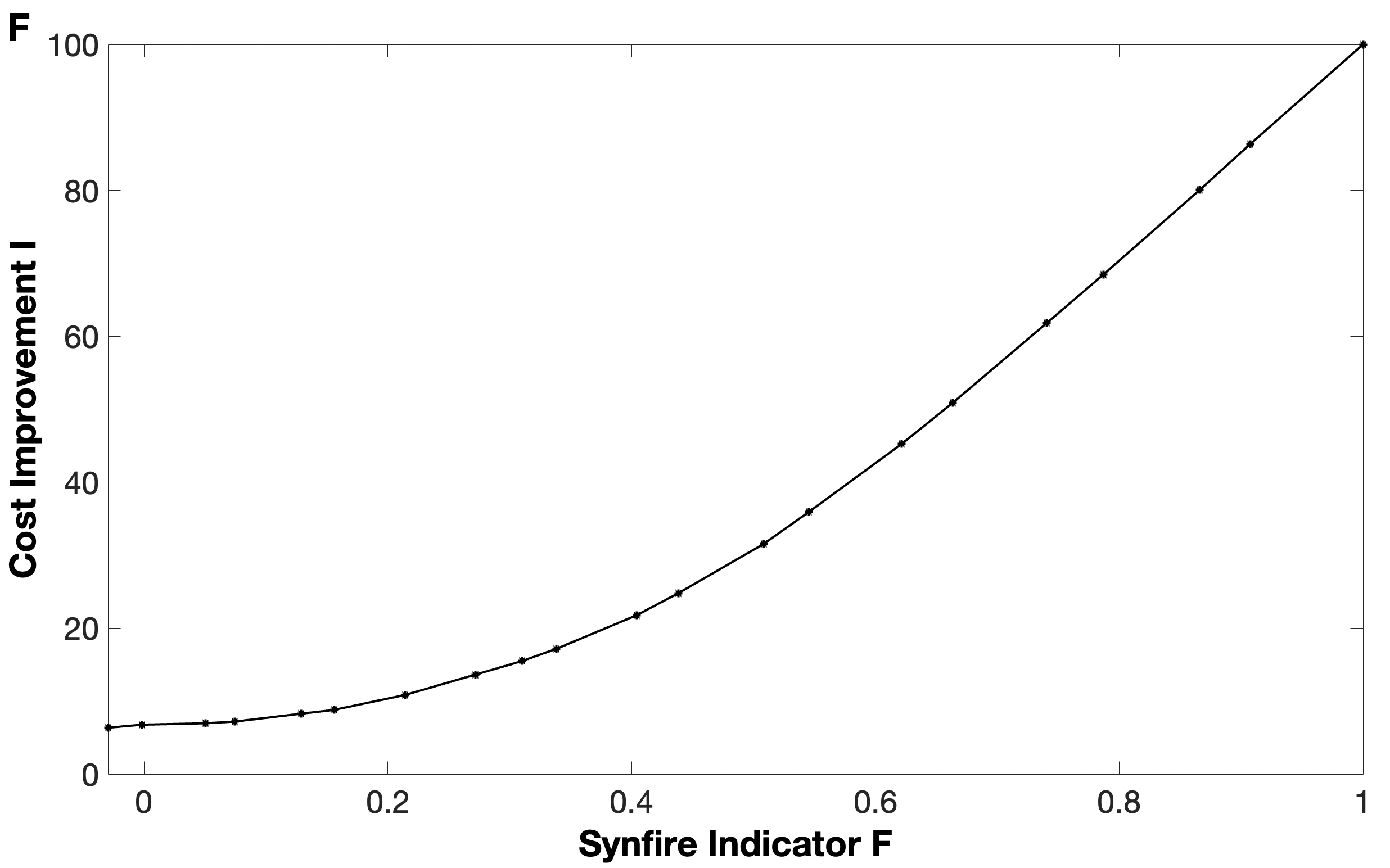}
    \end{subfigure}
    \caption{Simulations of experimental data: 3D-plots show the dependence of the Synfire Indicator $F$ (subplot A), SPIKE-Synchronization $C$ (subplot B) and the Cost improvement $I$ (subplot C) on the event completeness $P$ and the shuffle parameter $S$. Colored lines in A depict the crossings of the Synfire Indicator plane with horizontal planes corresponding to constant $F$-values, while in B and C they mark the values of $C$ and $I$, respectively, obtained for these parameter combinations of $P$ and $S$. In all three plots a thick black line marks the values obtained for SPIKE-synchronization $C$ equal to one. Subplot D depicts the values of event completeness and shuffle that yield the constant values of $F$ in subplot A and the maximum value of $C$ in subplot B. Finally, we show the cost improvement $I$ versus SPIKE-synchronization $C$ for different values of $F$ (subplot E) and versus the Synfire Indicator $F$ for $C = 1$ (subplot F). High values of the SPIKE-synchronization and, to an even larger extent, the Synfire Indicator are crucial for an impactful latency correction.}
\label{fig:F8_ResSim2_RealData_Sims}
\end{figure*}}

As we have seen in Fig. \ref{fig:F5_ResExp_Data_Example1_099_M29_D9}, a typical dataset consists of a number of rather complete global events (with a more or less consistent order), a few quite incomplete events and some noisy background spikes. We simulate all of these in a very controlled manner by setting the following parameters: the number of spike trains $N$, the number of global events $E$, the average relative completeness of these events $P$ and the relative amount of background spikes $B$ (in units of the number of spikes in the events if all of these events were complete). Note that both $P$ and $B$ are related to the mixing parameter $x$ from Section \ref{ss:Results-Sim} ($P = 1 - x$ and $B = x$) but here these two variables can be chosen independently and we are able to investigate parameter values beyond the ones covered in Section \ref{ss:Results-Sim} (for example $B > 1$) or beyond the values found in our experimental data of Section \ref{ss:Results-Exp}. Our last parameter, the shuffle $S$, controls the consistency of the spike order within the global events. It denotes the relative fraction of spikes in each event that are shuffled. For $S = 0$ nothing changes and a synfire chain remains a synfire chain, for $S = 0.5$ a randomly selected half of the spikes of each event are shuffled, while for $S = 1$ the shuffle is performed among all the spike trains present in each event and any initial consistency in order is completely destroyed.
%
%
\begin{figure*}[!ht]
\includegraphics[width=\linewidth]{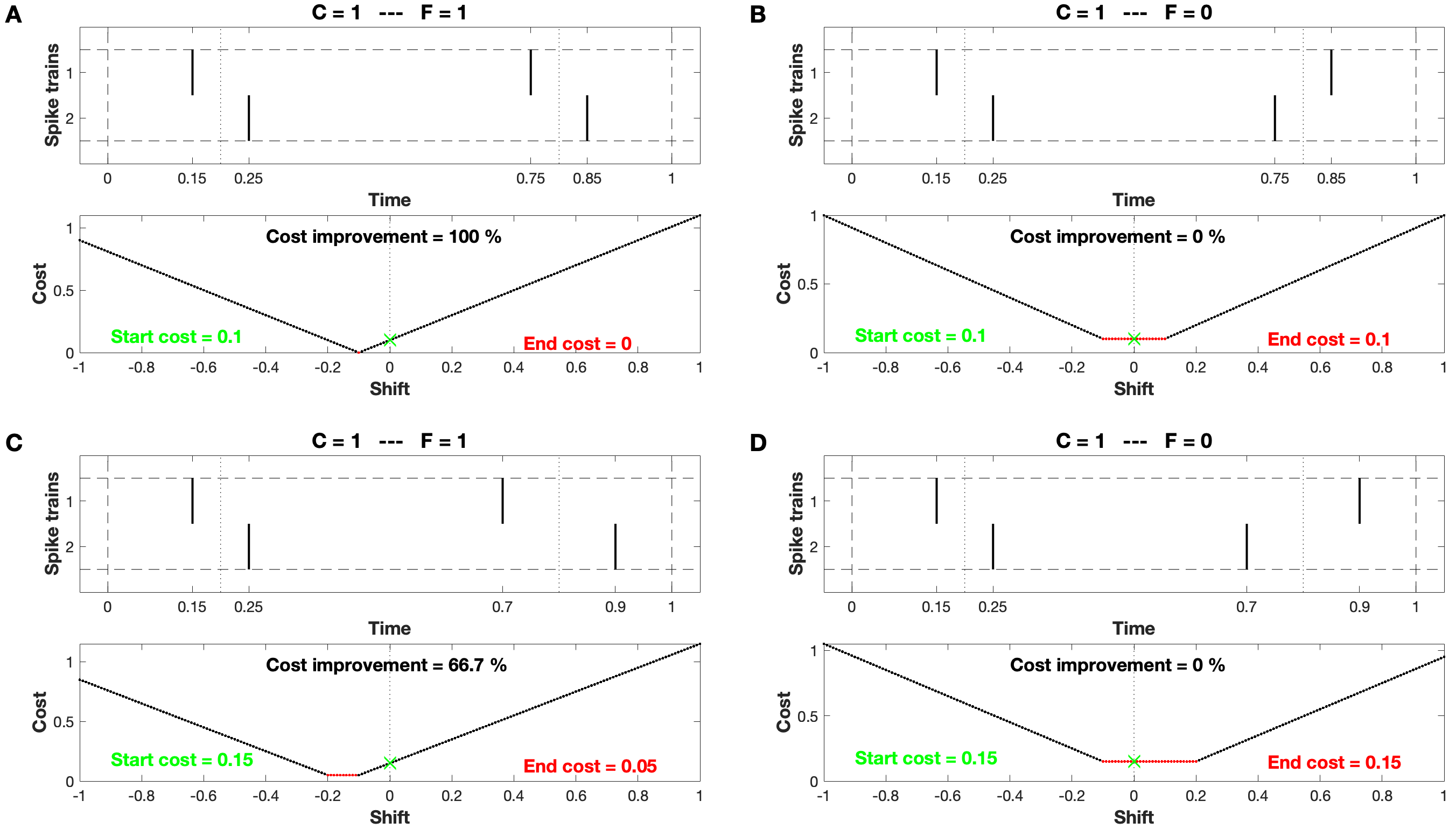}
\caption{Four arrangements of two spike trains with just two spikes each. In all four cases the upper plot shows two pairs of matching spikes ($C = 1$). Subplot A: Perfect synfire chain ($F = 1$). Subplot B: The two pairs exhibit opposite orders ($F = 0$). Subplot C: Consistent order, but variation in propagation velocity leading to different distances between the matched spikes. Subplot D: Opposite order and different propagation velocities. In the lower plots we show the cost value versus a potential `shift' of the second spike train with respect to the first. The start cost is the value at shift zero (marked by a green x), while the optimized end cost is the minimum value of this `cost function' (marked in red). While the value of the Synfire Indicator has a decisive influence on the success of the latency correction, the change in propagation velocity results in an offset in cost that persists and can thus be measured even after the latency correction.}
\label{fig:F9_ResExp_MinExamples-4Spikes}
\end{figure*}

With these parameters we can look more directly at the influence of SPIKE-synchronization $C$ (see Section \ref{ss:Methods-Spike-Matching-SimAnn}) and the Synfire Indicator $F$ (Section \ref{ss:Methods-SPIKE-order-Synfire-Indicator}) on the cost improvement and we do so in a way that the results on the one quantity are not disturbed or mediated by the other. The problem we have to overcome is that, while both the average completeness of events $P$ and the relative amount of background spikes $B$ are parameters that can be controlled easily, neither $C$ nor $F$ can be set directly. Fortunately, we can control the SPIKE-synchronization $C$ indirectly via $P$ and $B$ (starting from a perfect synfire chain both decreasing $P$ and increasing $B$ lead to smaller $C$). For simplicity, we here set $B$ to zero (no background noise), which maximizes the range of $C$ and $F$ values that we can cover. But we did confirm (results not shown) that for higher values of $B$ (including values larger than one) the results remain the same, the only difference is that for larger $B$ we are less able to reach the higher ranges of $C$ and $F$. 

So our two controlling parameters are the event completeness $P$ and the shuffle $S$ and on the left hand side of Fig. \ref{fig:F8_ResSim2_RealData_Sims} we display three 	3D-plots that show the dependence on these parameters of the Synfire Indicator $F$, SPIKE-synchronization $C$, and the cost improvement $I$, respectively (again all plots are averages over $100$ realizations). First, we note that the event completeness has an effect on both the Synfire Indicator and SPIKE-synchronization (the influence on $F$ is mediated via its upper limit $C$). The shuffle controls only the Synfire Indicator, whereas, as expected, SPIKE-synchronization is invariant to $S$. So a maximum value of $P$ leads immediately to a maximum of $C$, while $F$ is maximal if and only if events are both complete and consistently ordered (no shuffle). The same holds true for the cost improvement but while the decreases for $F$ and $C$ are (seemingly) linear, for $I$ the dependency on both $P$ and $S$ is non-linear.

Next, we use these 3D-plots to isolate the dependence of the cost improvement on SPIKE-synchronization and on the Synfire indicator. For $C$ we avoid mediation through $F$ by keeping $F$ constant. Thus we first identify in Fig. \ref{fig:F8_ResSim2_RealData_Sims}A the crossing of the `Synfire Indicator vs. event completeness and shuffle' plane with $11$ different horizontal planes corresponding to constant $F$-values from $0$ to $1$ (in steps of $0.1$). The projections onto the event completeness $P$ - shuffle $S$ - plane are shown in Fig. \ref{fig:F8_ResSim2_RealData_Sims}D and with these parameter combinations we can look up the corresponding values of SPIKE-synchronization in Fig. \ref{fig:F8_ResSim2_RealData_Sims}B and of the cost improvement in Fig. \ref{fig:F8_ResSim2_RealData_Sims}C. The resulting $I$ versus $C$ curves are shown in Fig. \ref{fig:F8_ResSim2_RealData_Sims}E. We find that for low values of $F$ the cost improvement is very small and there is hardly any dependence on SPIKE-synchronization. In this case the datasets are so noisy that not much can be gained by applying the latency correction algorithm. On the other hand, for high values of the Synfire Indicator the cost improvement increases considerably and so does the dependence on SPIKE-synchronization. It turns out that for constant $F$ the cost improvement actually decreases with $C$, a result which stands markedly in contrast to what we found in Section \ref{ss:Results-Exp}. There $I$ increased with $C$ which, as we learn here, was largely mediated by $F$. In fact, once $F$ is eliminated as an influencing factor, the dependency reverses. This new result can be explained as follows: For larger event completeness (and thus larger $C$) more shuffle is needed to keep $F$ constant (see Fig. \ref{fig:F8_ResSim2_RealData_Sims}D). The resulting datasets are more noisy and as before this noise keeps the cost improvement down. Reducing the Synfire Indicator via shuffle has a stronger effect on $I$ than reducing $F$ via the event completeness.

At last we turn our attention to the dependence on the Synfire Indicator $F$. This analysis is more straightforward, we can just use the event completeness $P$ to fix $C$ (according to Fig. \ref{fig:F8_ResSim2_RealData_Sims}B) and then vary the shuffle parameter $S$ to cover the whole range of $F$. We restrict ourselves to values of $P = 1$ (thus SPIKE synchronization $C = 1$) and look up the values of $F$ in Fig. \ref{fig:F8_ResSim2_RealData_Sims}A and of $I$ in Fig. \ref{fig:F8_ResSim2_RealData_Sims}C. The result is displayed in Fig. \ref{fig:F8_ResSim2_RealData_Sims}F which shows a very pronounced increase of the cost improvement with the Synfire Indicator. Lower values of $C$ (corresponding to parallel $P$-planes in Figs. \ref{fig:F8_ResSim2_RealData_Sims}A-C) yield similar results, just with a restricted $F$-range which also means that higher values of $I$ are no longer reached. The main result is that in all cases a high Synfire Indicator is essential for the proper functioning of the algorithm.

In our final figure we illustrate this importance of $F$ with minimal examples consisting of just two spike trains with two spikes each. As a representative of the noisy disturbances that are maintained by the latency correction, we also include the effect of different event lengths in our consideration. The upper plots of Fig. \ref{fig:F9_ResExp_MinExamples-4Spikes} show four possible arrangements of the four spikes. All of these arrangements consist of two perfectly matched spike pairs ($C$ = 1), the differences lie in the spike order and in the propagation velocity. The order in the two spike pairs is either consistent ($F = 1$, left subplots A and C) or inconsistent ($F = 0$, subplots B and D on the right), while the velocity is either constant (same interval within the two spike pairs, top subplots A and B) or varies (the second pair is more apart, subplots C and D at the bottom).

When we look at the four cost functions in the lower plots of Fig. \ref{fig:F9_ResExp_MinExamples-4Spikes} we see the main difference between the two kind of effects. Comparing left and right subplots we find that the inconsistency in order which is reflected in the value $F = 0$ for the synfire indicator results in a collapse of the cost improvement $I$. There is no systematic latency to correct and so the end cost equals the start cost. On the other hand, the different velocities in the two bottom arrangements lead to an offset in cost (in this case $0.05$, compare A with B and C with D) that is not affected by the latency correction, rather it is present before and after.

Taken together, these different arrangements illustrate the essential difference between the systematic delays that we correct with our algorithm and other non-systematic disturbances which are exactly the kind of deviations from synchrony that we wish to quantify once the correction has been achieved.

%
\section{Conclusions} \label{s:Conclusions}

In the quantification of synchrony, latency
is a systematic disturbance that first needs to be eliminated. While there have been some rate-based approaches for data with sufficiently large firing rates \cite{Nawrot03, Schneider06}, the problem of latency in sparse neuronal spike trains where the timing of individual spikes matters remained elusive. In the present study we address this issue and propose a latency correction algorithm that corrects any systematic delays but to maintains all other kinds of noisy disturbances in the data. It consists of two basic steps, spike matching and minimization of the distances between the matched spikes using simulated annealing.

The algorithm receives as input a set of spike trains (as typically shown in a rasterplot) and delivers three main outputs: the end cost, the shifts performed in order to get there, and the relative cost improvement. \tcr{The end cost (the minimal cost over the course of the simulated annealing) quantifies how well the spike trains in the dataset under investigation can be aligned.} The shifts (marked by arrows in Figs. \ref{F1_Synfire_chain} and \ref{F2_Simple_Example_Full_Plot}) provide information about the latencies that were present initially. \tcb{Finally, the relative cost improvement is a measure of the effect the algorithm has had in correcting the latency for this specific dataset.}

We validate the algorithm on controlled data simulated from scratch (Section \ref{ss:Results-Sim}), show its effectiveness in an experimental real life setting (global propagation patterns in the cortex of mice recorded via wide-field calcium imaging before and after stroke induction, Section \ref{ss:Results-Exp}) and use simulations of these experimental data (Section \ref{ss:Results-SimExp}) to identify the best conditions for its applicability to experimental datasets. In the first part we show that for mixings of a perfect synfire chain with Poisson spike trains decreasing the SPIKE-synchronization $C$ and the Synfire Indicator $F$ makes the cost improvement level off. On the real data we observe that the cost improvement is strongly positively correlated with the Synfire Indicator (and with SPIKE-synchronization as well, but much less). Finally, in the more systematic simulations of these real data we find again a pronounced increase of the cost improvement with the Synfire Indicator. On the other hand, for constant Synfire Indicators the cost improvement actually slightly decreases with SPIKE-synchronization (because more shuffle is needed to keep the Synfire Indicator constant and this noise limits the cost improvement).

Overall, we have accumulated evidence that the algorithm functions best for sparse data with well defined global events (as manifested by high values of $C$) and a consistent order within these events (corresponding to elevated values of $F$). But in particular in Section  \ref{ss:Results-SimExp} we have seen that these two quantities are not equally important. Clearly the one fundamental \textit{criterion} for a meaningful application of our latency correction algorithm is a \textit{high value of the Synfire Indicator $F$}. SPIKE-synchronization $C$ is not decisive in itself, but it is still relevant as a mediator: Without a reasonably high SPIKE-Synchronization there are not enough coincident spike pairs to estimate the latency within the spike trains. In addition, as we have mentioned repeatedly, SPIKE-Synchronization acts as an upper limit to the Synfire Indicator. Thus, a large value of SPIKE-Synchronization is a necessary condition, but as the minimum examples on the right hand side of Fig. \ref{fig:F9_ResExp_MinExamples-4Spikes} demonstrate, it is not a sufficient condition. If the spikes do not exhibit a high degree of consistency in order (a large $F$), the algorithm does not have enough systematic latency to work with. On the other hand, a high value of Synfire Indicator guarantees that SPIKE-Synchronization is large as well.

Nevertheless, it is important to stress that the actual value of the cost improvement also depends crucially on cost offset effects caused by noisy disturbances. What the algorithm does is correct constant systematic delays, any other disturbances are left unaffected. Thus the example for which the algorithm works best is a perfect synfire chain (as shown in Fig. \ref{fig:F3_ResSim1_Mixing-3Examples}A both $C$ and $F$ attain their maximum value of one). This is the only dataset that exhibits constant systematic delays but not any of the other potential sources of noise such as unreliability (missing spikes in the global events), jitter (noisy spike shifts), or background noise (extra spikes). These, together with other disturbances like different durations of global events (variation in propagation velocity) or changing intervals between subsequent spikes within the global events (non-monotonic propagation), make the correction more difficult and accordingly decrease the cost improvement. However, these are also exactly the deviations from synchrony in the dataset that we would like to quantify in the first place. We can easily do so by means of the cost offset that still remains even after the correction of the systematic delays has been performed. For an illustration of this point compare subplots A and C in Fig. \ref{fig:F9_ResExp_MinExamples-4Spikes}. These two examples both have maximum SPIKE-synchronization and Synfire Indicator, but the change in propagation speed in subplot C causes a cost offset that persists after the correction. For the algorithm this means a reduction of the cost improvement, but for our synchrony analysis it is just an indication of the noisiness in the dataset (and this is independent of whether a latency correction has been performed or not).
 
At the other extreme there are clearly some datasets where the algorithm does not work, in the sense that no significant improvement can be achieved. The most obvious case are datasets where there is no significant systematic latency at all, which means there is actually nothing to correct. One example is when the dataset is already perfectly synchronous. In this case $C$ is one and $F$ is zero. In most other such cases the data are rather disordered with very low values for both $C$ and $F$. A prominent example, Poisson spike trains, is shown in Fig. \ref{fig:F3_ResSim1_Mixing-3Examples}C.  

Another feature in the data that would create problems for the algorithm are global events that overlap and are thus difficult to entangle. More specifically, whenever the interval between two successive events is less than twice the propagation time within an event, according to the coincidence criterion of Eq. \ref{eq:Coincidence-MaxDist} there will be matchings between spikes from different events. Such mismatches would be indicated by a decrease in the value of SPIKE-synchronization $C$. Fortunately for us, many repetitive propagation phenomena in neuroscience as well in other fields (the algorithm is universal and could be applied to any type of discrete data) exhibit ratios of characteristic time scales that fulfill the coincidence criterion. For example, the duration of an epileptic seizure is usually much shorter than the interval between two successive seizures. Similarly, in meteorological data (an example outside of neuroscience) the time it takes a storm front to cross a specific region is typically much smaller than the time to the next storm.

A final rather general caveat that can be relevant under certain circumstances is that the latency we are referring to in this study is the latency from the point of view of the experimenter. However, sometimes (for example in case the dataset under consideration is derived from a neural network) it might be worth considering that every node in the network will have a different perspective of the same network activity which will depend on the array of propagation delays from each neuron to this ``observing" neuron.

We can identify three different areas of future directions. First, for the algorithm itself we envisage a way to get over some of these limits mentioned above, in particular the entanglement of overlapping successive events. The basic idea is to focus on the non-overlapping parts (the neighboring spike trains) and to restrict the definition of the cost function on the corresponding diagonals closest to the main diagonal thus disregarding the matrix elements disturbed by the overlap. After this modified latency correction has been carried out the correct matching of spikes can be performed thereby disentangling the overlapping events. This avenue has some complications as well as wider implications and will therefore be pursued in a forthcoming study.

Second, concerning the underlying aim of estimating synchrony, instead of using just the cost function itself (the average of the spike time difference matrix) as a measure of spike train synchrony, one could evaluate before but in particular after the correction more sophisticated and comprehensive measures of spike train similarity such as the time-scale independent ISI- \cite{Kreuz07c} or SPIKE-distances \cite{Kreuz13} or the time-scale dependent Victor-Purpura \cite{Victor96} or van Rossum \cite{VanRossum01} distances.

Finally, the most important point regards the experimental calcium data in mice before and after stroke that we analyzed. Together with our previous findings on the lower duration and increased smoothness \cite{Cecchini21}, the results of Fig. \ref{fig:F6_ResExp_RealData_Stats1} suggest that the propagation of cortical activation shows a faster, more coherent and linear pattern in the Combined group. We will follow up on these results and evaluate the potential of both the end cost value (Fig. \ref{fig:F6_ResExp_RealData_Stats1}) and the cost improvement (Fig. \ref{fig:F7_ResExp_RealData_Stats2}) to serve as biomarkers that are able to uncover neural correlates not only of motor deficits caused by stroke but also of functional recovery during the various rehabilitation paradigms. Such insights could pave the way towards more targeted post-stroke therapies.

The algorithm will be readily applicable for everyone since it will be implemented in three freely available software packages called SPIKY\footnote[1]{http://www.thomaskreuz.org/source-codes/SPIKY} (Matlab graphical user interface \cite{Kreuz15}), PySpike\footnote[2]{http://mariomulansky.github.io/PySpike} (Python library \cite{Mulansky16}) and cSPIKE\footnote[3]{http://www.thomaskreuz.org/source-codes/cSPIKE} (Matlab command line library with MEX-files). All of these software packages contain already the three symmetric measures of spike train synchrony, ISI-distance \cite{Kreuz07c, Kreuz09}, SPIKE-distance \cite{Kreuz11, Kreuz13}, SPIKE-synchronization \cite{Kreuz15} (see \cite{Satuvuori17} for generalized versions), the directional SPIKE-order \cite{Kreuz17} as well as source codes designed to find within a larger neuronal population the most discriminative subpopulation \cite{Satuvuori18b}.

\section*{Acknowledgement}

We thank Arturo Mariani for useful discussions and a careful reading of the manuscript.

This project has received funding from the H2020 EXCELLENT SCIENCE - European Research Council (ERC) under grant agreement ID n. 692943 BrainBIT and from the European Union’s Horizon 2020 Research and Innovation Programme under Grant Agreement No. 785907 (HBP SGA2) [Grant recipient: F.S.P.]. This research was supported by the EBRAINS research infrastructure, funded from the European Union’s Horizon 2020 Framework Programme for Research and Innovation under the Specific Grant Agreement No. 945539 (Human Brain Project SGA3) [Grant recipient: F.S.P.].


\bibliographystyle{elsarticle-num}

\end{document}